\documentclass{article}
\PassOptionsToPackage{hyphens}{url}
\usepackage{iclr2026_conference,times}

\usepackage{amsmath,amsfonts,bm}

\def\eqref#1{equation~\ref{#1}}

\def\1{\bm{1}}

\DeclareMathAlphabet{\mathsfit}{\encodingdefault}{\sfdefault}{m}{sl}
\SetMathAlphabet{\mathsfit}{bold}{\encodingdefault}{\sfdefault}{bx}{n}

\usepackage{hyperref}
\usepackage{float}
\usepackage{graphicx}
\graphicspath{{./images/}}
\usepackage{caption}
\usepackage[final]{microtype}
\microtypesetup{protrusion=true,expansion=true}
\usepackage{algorithm}
\usepackage{algorithmic}
\usepackage{booktabs}
\usepackage{array}
\usepackage{amsmath}
\usepackage{amssymb}
\usepackage{tabularx}
\usepackage{multirow}
\usepackage{pgfplots}
\pgfplotsset{compat=1.18}
\usepackage{placeins}
\usepackage{fancyvrb}
\usepackage{fvextra}
\DefineVerbatimEnvironment{promptbox}{Verbatim}{%
  fontsize=\scriptsize,
  frame=single,
  framesep=3pt,
  breaklines=true,
  breakanywhere=true,
  breaksymbolleft={},
  breakindent=0pt,
}

\newcommand{\scorepm}[2]{#1\,{\scriptsize$\pm$}\,#2}

\title{NexForge: Scaling Agent Capabilities through Requirement-Driven Task Synthesis for LLMs}

\author{Jiarong Zhao\textsuperscript{1}, Zhikai Lei\textsuperscript{3,*}, Zhiheng Xi\textsuperscript{2}, Rui Zheng\textsuperscript{3}, Hang Yan\textsuperscript{3}, \textbf{Jie Zhou}\textsuperscript{1,*}, \\\textbf{Qin Chen}\textsuperscript{1}, \textbf{Liang He}\textsuperscript{1} \\[4pt]
\textsuperscript{1}East China Normal University \quad \textsuperscript{2}Fudan University \quad \textsuperscript{3}Shanghai Qiji Zhifeng Co., Ltd \\[2pt]
\textsuperscript{*}Corresponding authors
}
\iclrfinalcopy
\begin{document}

\maketitle
\lhead{}

\begin{abstract}
Scaling executable agent training data for LLM post-training is bottlenecked by substrate-bound methods that tie task generation to predefined tools, repositories, or skill graphs: expanding coverage requires manual substrate engineering, each new domain demands a bespoke pipeline, and the resulting task distributions often reflect substrate biases rather than real-world demand. We introduce NexForge, a requirement-driven framework that takes high-level capability requirements as input and synthesizes diverse, executable agent tasks and expert trajectories for SFT. NexForge first investigates real-world demand to construct scenarios and task profiles, then performs distribution-aware compilation to generate task directives. For each directive, NexForge automatically retrieves or constructs the required files, dependencies, and runtime configurations, and finally collects expert rollouts to produce training trajectories. Without domain-specific infrastructure, NexForge produces 3.6K terminal and 2K office tasks, improving Qwen3.5-35B-A3B Base from 22.5\% to 52.0\% on Terminal-Bench 2.0 and from 813 to 1338 Elo on GDPval; scaling further to 43.2K terminal tasks yields 58.4\%, on par with Claude Opus 4.6 equipped with Claude Code. Scaled further, NexForge-synthesized data contributes to the training of Nex-N2, a family of publicly available agent models that lift Qwen3.5-397B-A17B to 75.3\% on Terminal-Bench 2.1 and to 1585 Elo on GDPval---achieving state-of-the-art open-source performance and surpassing several frontier proprietary systems. Nex-N2 models are available at \url{https://nex.sii.edu.cn/}.
\end{abstract}

\begin{figure}[!b]
\centering
\includegraphics[width=\textwidth]{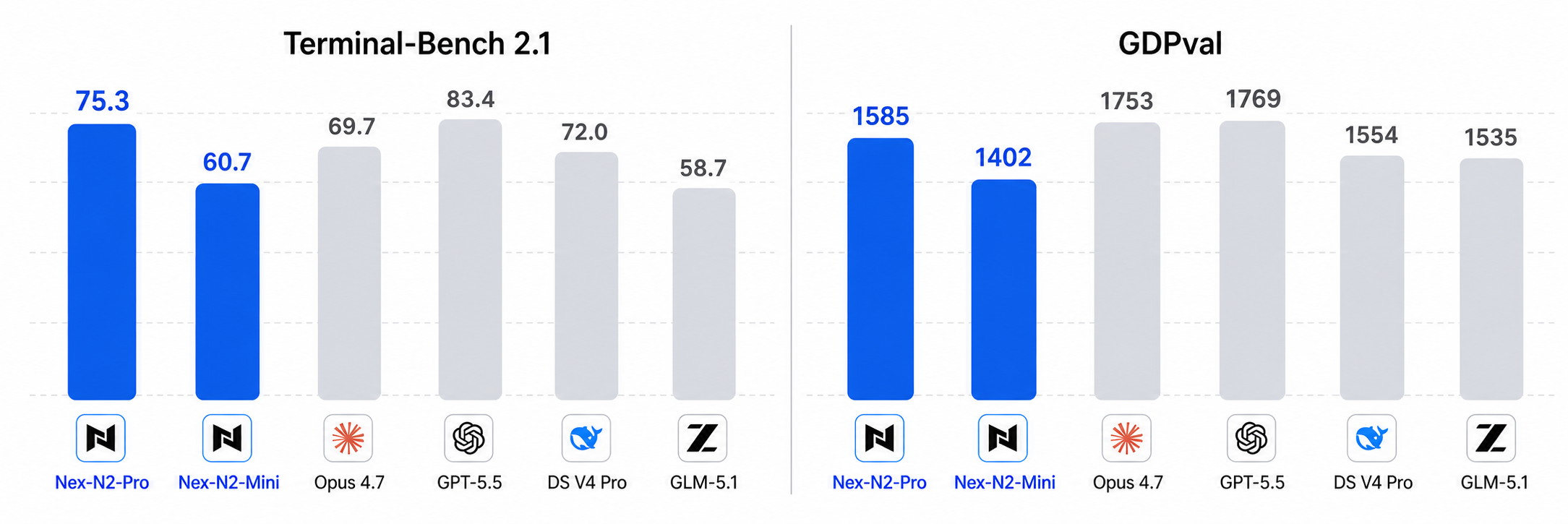}
\caption{\textbf{Nex-N2 performance on Terminal-Bench 2.1 and GDPval via NexForge data scaling.}}
\label{fig:teaser}
\end{figure}

\section{Introduction}
\label{sec:introduction}

Building capable LLM agents for real-world tasks is bottlenecked by the scarcity of diverse training data. Post-training these agents requires not only task descriptions, but also grounded materials, executable environments, and long-horizon interaction trajectories. Producing such data at scale remains labor-intensive: practitioners must design representative tasks, collect appropriate materials, prepare runtime environments, and package everything into runnable workspaces---an expensive, domain-specific process that resists efficient scaling.

\begin{figure}[t]
\centering
\includegraphics[width=0.7\textwidth]{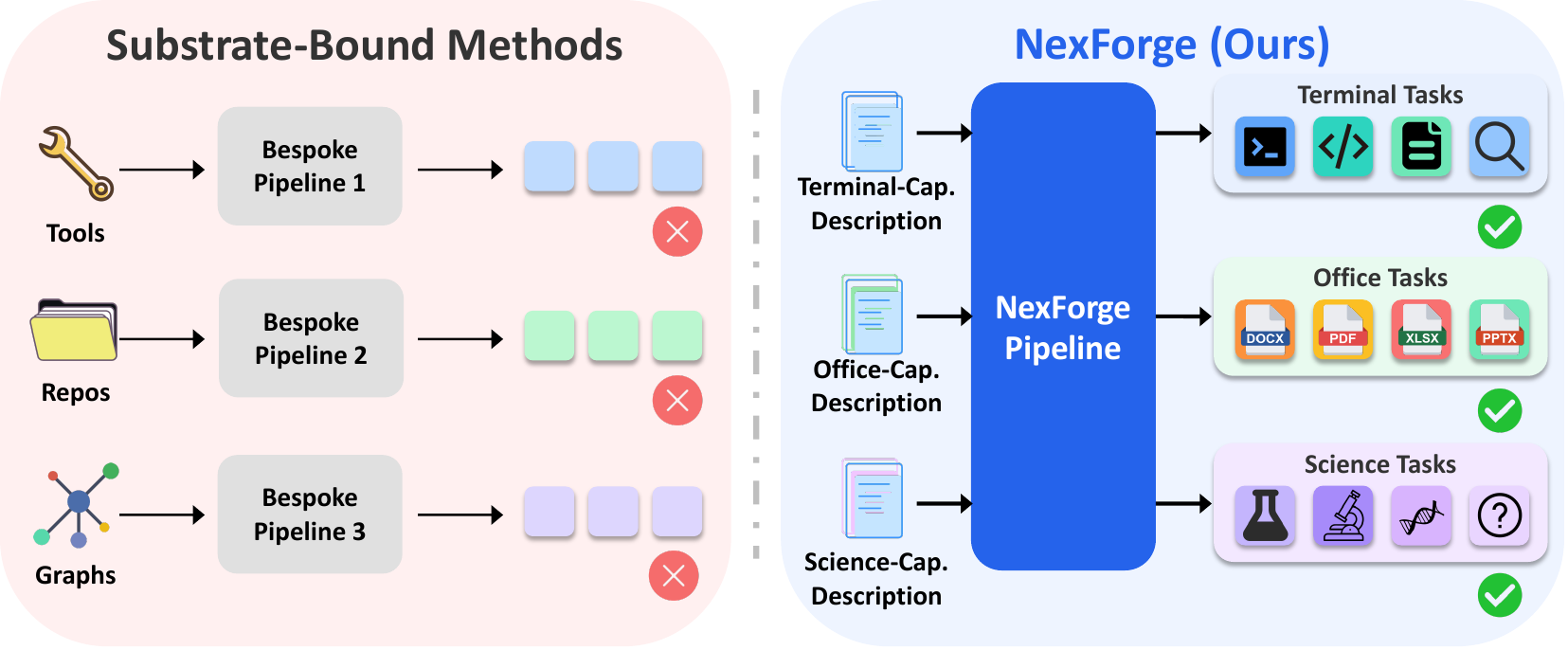}
\vspace{-3mm}
\caption{\textbf{Substrate-bound Methods vs.\ NexForge} Existing methods (\textcolor{red!70!black}{left}) tie task generation to predefined substrates---tools, repositories, or skill graphs---whose coverage and scale are bounded by the substrate (\textcolor{red!70!black}{$\times$}). NexForge (\textcolor{blue!80!black}{right}) decouples task generation from substrates via a unified core driven by capability requirements, enabling the same pipeline to scale freely across domains (\textcolor{green!60!black}{$\checkmark$}).}
\label{fig:motivation}
\vspace{-4mm}
\end{figure}

Recent task-synthesis methods generate executable tasks tied to predefined repositories, tools, skills, or execution traces~\citep{dong2026agentworld,wang2026agentworldmodel,chen2026dive,xie2025agentsynth,fan2026skillsynth,cheng2026terminalworld,yang2025swesmith,jain2025r2egym}. As illustrated in Figure~\ref{fig:motivation} (left), these approaches share three limitations: scale is bounded by the input substrate, each domain demands bespoke infrastructure, and the substrate implicitly determines the task distribution, which reflects substrate biases rather than real-world demand.

We argue that scaling agent training data requires decoupling task generation from predefined substrates. Environment synthesis should instead begin with the user's capability requirement: the system first identifies which task types are representative of the desired capability and how frequently they occur, and only then determines which materials and runtimes are needed. This requirement-driven formulation separates \emph{what an agent should practice} from \emph{how that practice is made executable}, enabling the same pipeline to scale across domains without building new domain-specific infrastructure for each target.

To this end, we introduce NexForge, a requirement-driven framework that synthesizes diverse, executable agent tasks and expert trajectories for SFT, all from high-level capability requirements. NexForge investigates real-world demand to construct scenarios and task profiles, compiles distribution-aware task directives by composing sampled task forms with scenarios, and instantiates each directive into an executable workspace with expert trajectories. We evaluate NexForge on two disparate capabilities: 3.6K tasks for terminal operations improve Qwen3.5-35B-A3B Base from 22.5\% to 52.0\% on Terminal-Bench 2.0; 2K tasks for office work improve the same model from 813 to 1338 Elo on GDPval, both datasets synthesized without any domain-specific engineering effort. Scaling up the training data further enables Nex-N2 to reach state-of-the-art open-source performance on both capabilities, as shown in Figure~\ref{fig:teaser}.

Beyond controlled experiments, NexForge serves as a core data engine behind the Nex-N2 model family~\citep{nexn2website}, a series of publicly available agent models trained on scaled-up NexForge corpora. NexForge-synthesized tasks span terminal operations, office productivity, and long-horizon agentic reasoning (OpenClaw), providing the diverse, executable training signals that underpin Nex-N2's capabilities across these domains. Nex-N2-Pro (Qwen3.5-397B-A17B) achieves 75.3\% on Terminal-Bench 2.1 and 1585 Elo on GDPval, competitive with frontier proprietary systems including GPT-5.5 and Claude Opus 4.7; Nex-N2-Mini (Qwen3.5-35B-A3B Base), trained on a smaller data budget, reaches 60.7\% and 1402 Elo respectively---already surpassing the controlled Terminal-3.6K run. The gap between the controlled runs and the Nex-N2 models demonstrates that requirement-driven data synthesis scales effectively from research-stage corpora to production-grade LLM post-training.

Our contributions are threefold:
\begin{itemize}
\item We formulate environment scaling for agent post-training as a \emph{requirement-driven scaling problem}, identifying three bottlenecks of substrate-bound synthesis: input-bounded scale, high transfer cost, and substrate-biased distributions.
\item We propose NexForge, an end-to-end framework that investigates real-world demand, instantiates executable training environments, and collects expert trajectories, without domain-specific infrastructure.
\item We demonstrate that the same framework generates effective training data for both terminal and office capabilities, substantially improving downstream agent performance across distinct domains.
\end{itemize}

\section{Related Works}

\paragraph{Instruction and task-distribution synthesis.}
Scalable data synthesis often decomposes a broad capability space before generating individual examples. Self-Instruct~\citep{wang2023selfinstruct} bootstraps instruction-following data from a small seed set, while Evol-Instruct~\citep{xu2024wizardlm} evolves instructions through iterative complexity augmentation. GLAN~\citep{li2024glan} organizes generation through a hierarchical taxonomy. TRouter~\citep{liu2026trouter} controls synthetic examples using explicit task profiles, and Benchmark Agent~\citep{xiong2026benchmarkagent} decomposes evaluation goals into structured benchmark specifications. These studies show that explicit distribution modeling improves coverage compared to unconstrained generation, but they primarily synthesize text instructions or evaluation benchmarks rather than complete executable environments. Extending these approaches to agent training would additionally require grounding each task with materials, dependencies, and executable runtime environments.

\paragraph{Agent task and environment synthesis.}
Recent work has explored executable agent-data construction through simulated environments and tool ecosystems~\citep{dong2026agentworld,wang2026agentworldmodel}, execution traces and subtask composition~\citep{chen2026dive,xie2025agentsynth,shi2025taskcraft}, and domain-specific repositories and skills~\citep{fan2026skillsynth,cheng2026terminalworld,yang2025swesmith,jain2025r2egym}. Agent-World~\citep{dong2026agentworld} and Agent World Model~\citep{wang2026agentworldmodel} synthesize stateful tool environments, while DIVE~\citep{chen2026dive} derives grounded and verifiable tasks from execution evidence. AgentSynth~\citep{xie2025agentsynth} and TaskCraft~\citep{shi2025taskcraft} construct complex tasks through subtask composition and difficulty control. For terminal agents, SkillSynth~\citep{fan2026skillsynth} generates tasks from structured skill specifications, Terminal-World~\citep{cheng2026terminalworld} builds environments around reusable agent skills, CLI-Universe~\citep{hua2026cliuniverse} constructs terminal tasks from a predefined capability taxonomy, and TerminalTraj~\citep{wu2026terminaltraj}, Nemotron-Terminal~\citep{pi2026terminaldata}, LiteCoder-Terminal~\citep{peng2026litecoderterminal}, and OpenThoughts-Agent~\citep{raoof2026openthoughtsagent} collect or curate terminal interaction trajectories and training corpora. In software engineering, SWE-smith~\citep{yang2025swesmith} and R2E-Gym~\citep{jain2025r2egym} generate executable tasks from repositories, tests, and issue-like workflows.

These methods establish the importance of executable environments but share the substrate-bound limitations discussed before. NexForge addresses all three limitations by starting from high-level capability requirements.

\section{Method}

\begin{figure}[t]
\centering
\includegraphics[width=0.99\textwidth]{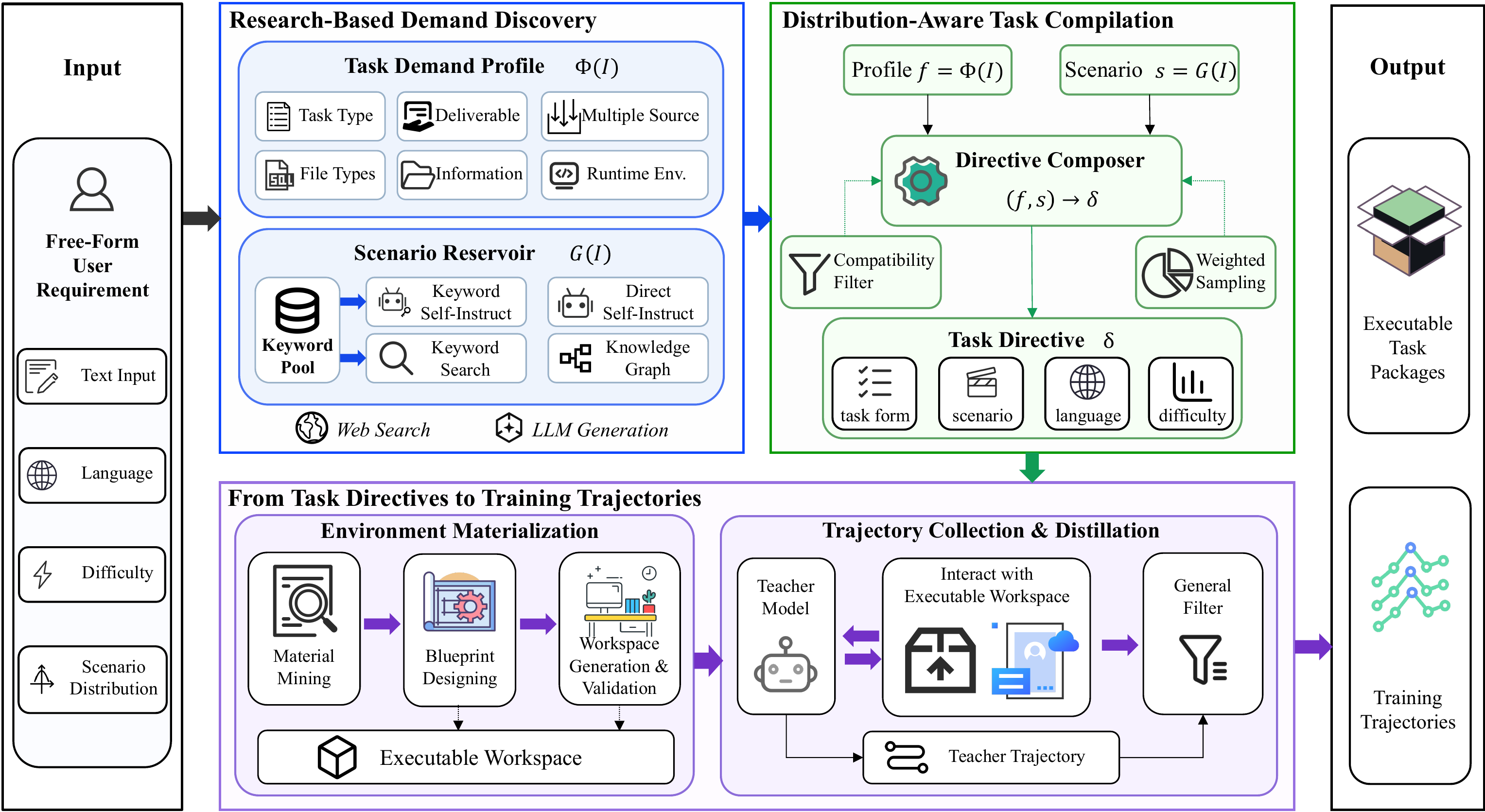}
\caption{\textbf{Overview of NexForge.} Given a high-level user requirement, NexForge first investigates real-world demand to construct a task demand profile and scenario reservoir, then composes distribution-aware task directives across scenarios. Each directive is subsequently instantiated into an executable workspace, where teacher interactions are collected and distilled into training trajectories.}
\label{fig:framework}
\end{figure}

NexForge takes a high-level user requirement and produces diverse executable workspaces and interaction trajectories for agent post-training. As illustrated in Figure~\ref{fig:framework}, the framework proceeds through three stages---demand discovery, task compilation, and trajectory generation---whose pipeline determines \emph{what capabilities the agent should acquire} before constructing \emph{the materials and runtime environments that make the practice executable}, enabling the same architecture to scale across agent capability targets.

\subsection{Problem Formulation}

NexForge takes as input a high-level capability requirement $I$, a requested task count $N$, and optional batch-level constraints $B$, such as language and difficulty budgets. It outputs a collection
\begin{equation}
    \mathcal{T}={\tau_1,\ldots,\tau_N},
\end{equation}
where each $\tau_i$ is an executable task package containing a task instruction, grounded materials, workspace files, software dependencies, and a runtime configuration. These packages are designed for trajectory collection rather than benchmarking and therefore do not require manually authored reference answers or task-specific verifiers.

We represent each task with task form $f_i$ and scenario $s_i$:
\begin{equation}
    \tau_i \leftarrow (f_i,s_i).
\end{equation}

The task form $f_i=(t_i,d_i,\sigma_i,e_i,\ell_i,h_i)$
specifies the primary task type $t_i$, expected deliverable $d_i$, source strategy $\sigma_i$, runtime environment $e_i$, language $\ell_i$, and difficulty $h_i$. The scenario describes the concrete context in which the task occurs, such as an organization, professional role, workflow, software system, or operational event. This decomposition enables explicit control over the corpus-level task distribution while grounding each task in a realistic context.

\subsection{Research-Based Demand Discovery}

Given the requirement $I$, NexForge first investigates what the requested capability involves in practice. A web-enabled research agent collects evidence from professional workflows, technical documentation, role descriptions, public examples, representative artifacts, and other domain-relevant sources. The collected evidence is organized into a weighted \emph{task demand profile} and a diverse \emph{scenario reservoir}.

\paragraph{Task demand profile.}

The task demand profile $\Phi(I)$ defines the intended distribution over the dimensions of the task form:
\begin{equation}
\Phi(I)=
\{\Phi_t,\Phi_d,\Phi_\sigma,\Phi_e,\Phi_\ell,\Phi_h\},
\end{equation}
where each $\Phi_k$ is a weighted categorical distribution. Candidate options are derived from research evidence and model knowledge, and are subsequently reviewed by an LLM to merge near-duplicates, remove ambiguous categories, and improve mutual exclusivity. Evidence-derived weights govern the relative sampling proportion of each option for the requested capability.

The profile specifies the composition of the generated corpus without committing to repositories, documents, or source files. It therefore separates the intended work distribution from the materials used to instantiate tasks.

\paragraph{Scenario reservoir.}

The scenario reservoir $\mathcal{G}(I)$ contains concrete scenario guides (\emph{scenarios} for short) that describe working contexts in which task forms can be instantiated. NexForge populates it through four mechanisms: (1)~\emph{keyword-conditioned self-instruct}, which samples keywords from a domain keyword pool and prompts the model to generate scenarios; (2)~\emph{keyword-conditioned research}, which uses sampled keywords as web-search queries to discover real-world scenarios; (3)~\emph{self-instruct}, which directly prompts the model to generate scenarios, sampling from already accepted scenarios as few-shot examples to encourage novelty; and (4)~\emph{knowledge-graph expansion}, which iteratively expands sub-scenarios from the input requirement. The domain keyword pool is generated via web research or model knowledge. The relative sampling proportion of each mechanism can be configured through the batch-level constraints $B$.

Scenarios intentionally describe only the working context and omit task type, deliverable, source strategy, and runtime. This prevents scenarios from independently determining the task distribution. All candidate scenarios undergo exact-duplicate removal and embedding-based semantic novelty filtering against previously accepted scenarios, producing a broad but non-redundant scenario reservoir.

\subsection{Distribution-Aware Task Compilation}

NexForge composes the task-form profile and scenario reservoir into directives. For each scenario $g_j\in\mathcal{G}(I)$, the composer first expands it into a detailed scenario description and then sequentially filters the profile options according to their compatibility with the scenario and the previously selected fields.

Let $K={t,d,\sigma,e}$ denote the ordered task-form dimensions that require compatibility decisions. For each dimension $k\in K$, the composer obtains a compatible candidate set
$$
C_{j,k}=F_\theta(I,g_j,k,\Phi_k,S_{j,<k}),
$$
where $S_{j,<k}$ contains fields selected in earlier steps. A field is then sampled according to the profile weights restricted to the compatible set:
$$
f_{j,k}
\sim
\operatorname{Normalize}
\left(
\Phi_k\vert_{C_{j,k}}
\right).
$$
Language and difficulty are sampled according to the corresponding profile distributions and batch constraints.

The resulting directive is
$$
\delta_j=
\langle
g_j,f_j,A_j,\ell_j,h_j,R_j
\rangle,
$$
where $A_j$ records the compatible candidate sets and $R_j$ stores the composition rationale. The directive determines the intended work before any repository or document is selected, constraining downstream stages to find or construct materials that realize the sampled task. The compatibility filter ensures that the resulting directive is internally consistent; its effect on directive quality is evaluated in the ablation studies.

\subsection{Agent Post-Training Trajectory Generation}

A task directive specifies the intended work but is not yet executable. NexForge instantiates each directive into an executable workspace and subsequently collects teacher interactions and produces post-training trajectories.

\paragraph{Environment instantiation.}

Given a task directive $\delta_j$, NexForge instantiates the intended task through three successive steps: material mining, blueprint planning, and workspace generation and validation. A research agent first identifies realistic workflows and collects the resources required by the directive, including public repositories, technical documents, datasets, spreadsheets, configuration files, and other domain-relevant artifacts. When suitable external resources are unavailable, it specifies the local materials that should be generated. A planning agent then inspects the collected resources and produces a structured blueprint defining the task objective, input materials, expected deliverable, workspace organization, required transformations, software dependencies, and runtime constraints. Based on this blueprint, an implementation agent constructs the executable workspace by adapting repositories, retrieving public files, generating local artifacts, installing dependencies, and configuring an unprivileged CPU-only Docker environment. Automated checks verify material completeness, source-strategy consistency, dependency availability, and the absence of leaked solutions or target artifacts, and enforce runtime restrictions. This three-phase separation prevents material availability from altering the task intent.

\paragraph{Trajectory collection and filtering.}

Each workspace is assigned to a teacher model under a fixed interaction budget. The teacher model interacts with the executable environment using the available tools, producing a trajectory that records model responses, tool calls, environment observations, intermediate failures, and recovery behaviors. NexForge then converts the collected interactions into a standardized training format through task-independent trajectory cleaning, removing malformed, degenerate, or trivially short interactions. The framework does not require manually authored reference answers or task-specific success verifiers; therefore, incomplete but meaningful trajectories are retained when they contain useful long-horizon reasoning, tool-use patterns, or error-recovery signals.

\section{Datasets}
\label{sec:datasets}

We apply NexForge to two separately authored capability requirements, producing \textbf{Terminal-3.6K} for terminal agent post-training and \textbf{Office-2K} for office agent post-training. To study data scaling, we additionally produce Terminal-2K (the first 2{,}000 tasks of Terminal-3.6K), Terminal-43.2K, and Office-22K using the same pipeline.

\subsection{Task-Set Specifications}
\label{sec:dataset-specifications}

The terminal specification covers agentic command-line capabilities such as repository inspection, software building, configuration editing, data processing, system operation, and local validation, while the office specification includes agentic office-work capabilities such as spreadsheet analysis, document review, information aggregation, summarization, drafting, planning, and evidence-based recommendations. Both use a 50/50 Chinese--English split and balanced difficulty distributions. We choose the terminal corpus size (3,600 tasks with 3 rollouts each) to match the experimental design of SkillSynth~\citep{fan2026skillsynth}, enabling direct comparison at the same data budget; Office-2K contains 2,000 independently synthesized office tasks.\footnote{All synthesis stages use GPT-5.5.}

\subsection{Corpus Statistics and Rollouts}
\label{sec:dataset-statistics}

Table~\ref{tab:dataset-statistics} summarizes corpus scale, trajectory complexity, task diversity, and grounding characteristics. Following SkillSynth~\citep{fan2026skillsynth}, we collect 3 teacher rollouts per task for the controlled corpora (Terminal-2K, Terminal-3.6K, and Office-2K); the larger-scale Terminal-43.2K and Office-22K use a single rollout per task. All synthesis agents and the composer are driven by GPT-5.5~\citep{openai2026gpt55}. After trajectory conversion and task-independent trajectory cleaning, this yields 8,521 trajectories for Terminal-3.6K and 5,940 for Office-2K (both originally collected at 3 per task), reflecting natural failure rates.

The two specifications induce clearly different data distributions. Terminal-3.6K realizes 24 task types and is primarily grounded through public repository adaptation, whereas Office-2K realizes 15 task types and more frequently relies on public documents, spreadsheets, and reports. The ten most frequent task signatures account for only 20.6\% and 15.1\% of the two corpora, respectively, indicating that scale is not obtained by repeatedly instantiating a small number of templates.

Full synthesis configurations, token statistics, and run provenance are provided in the supplementary material.

\begin{table}[t]
\caption{\textbf{Corpus statistics.} Task counts, trajectory scale, and ablation variants. \emph{w/o profile} removes task-form sampling; \emph{w/o scenario} replaces grounded scenarios with neutral seeds.}
\label{tab:dataset-statistics}
\centering
\small
\setlength{\tabcolsep}{5pt}
\renewcommand{\arraystretch}{1.15}
\begin{tabular}{@{}lcccc@{}}
\toprule
\textbf{Statistic}
& \textbf{Term.-3.6K}
& \textbf{Office-2K}
& \textbf{w/o profile}
& \textbf{w/o scenario} \\
\midrule
Executable tasks
& 3,600
& 2,000
& 2,000
& 2,000 \\
Converted trajectories
& 8,521
& 5,940
& 4,574
& 4,541 \\
Trajectories per task
& 2.37
& 2.97
& 2.29
& 2.27 \\
Avg.\ tokens per traj.
& 129K
& 107K
& 124K
& 154K \\
Avg.\ tool calls per traj.
& 123
& 54
& 116
& 133 \\
Chinese / English
& 50/50
& 50/50
& 50/50
& 50/50 \\
Difficulty allocation
& Balanced
& Balanced
& Balanced
& Balanced \\
Realized task types
& 24
& 15
& ${\sim}$3
& ${\sim}$16 \\
Top-10 signature cov.
& 20.6\%
& 15.1\%
& 78.5\%
& 24.9\% \\
Mean package size
& 10.9 MB
& 9.4 MB
& 5.8 MB
& 9.6 MB \\
\bottomrule
\end{tabular}
\end{table}

\subsection{Comparison with Existing Agent Datasets}
\label{sec:dataset-comparison}

Table~\ref{tab:dataset-comparison} compares NexForge with representative agent-task synthesis pipelines. Existing methods typically start from predefined tools, repositories, atomic tasks, or designed skill structures, each requiring domain-specific infrastructure that limits scale and transfer. NexForge instead takes a high-level user requirement as input and profiles real-world demand before constructing materials and environments, separating corpus-level task distribution from grounding resources.

\begin{table}[t]
\caption{\textbf{Dataset comparison.} Scale reports tasks / trajectories as described in the respective publications. Columns: free-form requirement (Req.), demand research (Res.), corpus distribution control (Dist.), automatic environment instantiation (Env.), and cross-capability transfer (Cap.). $\triangle$: partial support.}
\label{tab:dataset-comparison}
\centering
\setlength{\tabcolsep}{4pt}
\resizebox{\textwidth}{!}{%
\begin{tabular}{@{}lccccccc@{}}
\toprule
\textbf{Dataset}
& \textbf{Synthesis Basis}
& \textbf{Scale}
& \textbf{Free-form Req.}
& \textbf{Demand Res.}
& \textbf{Corpus Dist.}
& \textbf{Auto Env.}
& \textbf{Cross-cap.} \\
\midrule
AgentSynth
& Composable subtasks
& 6K / 6K
& -- & -- & $\triangle$ & -- & $\triangle$ \\
TaskCraft
& Atomic tasks, tools
& 3.2K / 3.2K
& -- & -- & $\triangle$ & -- & $\triangle$ \\
DIVE
& Executed tool traces
& 114K / 48K
& -- & -- & $\triangle$ & -- & $\triangle$ \\
R2E-Gym
& Repositories, commits
& 2K / 3.3K
& -- & -- & -- & $\checkmark$ & -- \\
SWE-smith
& Repositories, tests
& 8.7K / 5K
& -- & -- & -- & $\checkmark$ & -- \\
SkillSynth
& Skill graph
& 3.6K / 10.7K
& -- & -- & $\triangle$ & $\checkmark$ & -- \\
Terminal-World
& Agent skill graphs
& 5.7K / 5.7K
& -- & -- & $\triangle$ & $\checkmark$ & -- \\
CLI-Universe
& Capability taxonomy
& 6K / 6K
& -- & $\checkmark$ & $\triangle$ & $\checkmark$ & -- \\
\midrule
\textbf{NexForge}
& \textbf{User requirement}
& \textbf{5.6K / 14.5K}
& $\checkmark$
& $\checkmark$
& $\checkmark$
& $\checkmark$
& $\checkmark$ \\
\bottomrule
\end{tabular}%
}
\end{table}

\section{Experiments}
\subsection{Experimental Setup}

\paragraph{Evaluation.} We evaluate terminal and office capabilities on Terminal-Bench 2.0~\cite{merrill2026terminalbench,terminalbenchleaderboard2026} and GDPval~\cite{patwardhan2025gdpval}, respectively; both benchmarks are used exclusively for downstream evaluation and are not involved in data construction. All models use the NexAU harness with consistent tools and inference settings. The baseline results are from published papers or public leaderboards and may use different scaffolds. We compare our models with frontier proprietary models and terminal-specialized baselines built on Qwen3-32B, including LiteCoder-Terminal \cite{peng2026litecoderterminal}, TerminalTraj \cite{wu2026terminaltraj}, OpenThinkerAgent~\citep{raoof2026openthoughtsagent}, Nemotron-Terminal \cite{pi2026terminaldata}, SkillSynth \cite{fan2026skillsynth}, CLI-Universe \cite{hua2026cliuniverse}, and Terminal-World \cite{cheng2026terminalworld} among others.

\paragraph{Implementation Details.} We perform full-parameter SFT on Qwen3-32B and Qwen3.5-35B-A3B Base~\citep{qwen2025qwen3,qwen2026qwen35} for five epochs with a maximum context length of 262K, packed bfloat16 sequences, a global batch size of 64, a learning rate of $10^{-5}$ with cosine decay and 5\% warmup. Trajectories are collected by a DeepSeek-V4-Pro~\citep{deepseek2026v4} teacher agent using the NexAU scaffold with file editing, bash execution, and web search tools, running three rollouts per task in the controlled settings and one rollout per task in the scaled runs. All Terminal-Bench 2.0 results are reported as mean $\pm$ standard deviation over four independent evaluations.

\newsavebox{\tbresultsbaselinebox}
\begin{table}[!t]
\caption{\textbf{Terminal-Bench 2.0 pass@1 accuracy.} Left: baseline models; right: terminal-specialized 32B models. CLI-Universe-32B (DS) uses a DeepSeek teacher. NexAU rows are our controlled runs.}
\label{tab:tb-results}
\centering
\small
\setlength{\tabcolsep}{2pt}
\sbox{\tbresultsbaselinebox}{%
\begin{minipage}[t]{0.485\textwidth}
\begin{tabular*}{\linewidth}[t]{@{}>{\raggedright\arraybackslash}p{0.38\linewidth}@{\extracolsep{\fill}}>{\centering\arraybackslash}p{0.12\linewidth}>{\centering\arraybackslash}p{0.25\linewidth}>{\centering\arraybackslash}p{0.20\linewidth}@{}}
\multicolumn{4}{c}{\textbf{Baselines}} \\
\cmidrule{1-4}
\textbf{Model} & \textbf{Params} & \textbf{Scaffold} & \textbf{TB 2.0} \\
\midrule
Gemini 3.1 Pro & -- & TongAgents & \scorepm{80.2}{2.6} \\
DeepSeek-V4-Pro & 1.6T & official & 67.9 \\
Claude Opus 4.6 & -- & Claude Code & \scorepm{58.0}{2.9} \\
MiniMax M2.7 & 230B & official & 57.0 \\
GPT-5.2 & -- & \mbox{Terminus 2} & \scorepm{54.0}{2.9} \\
GLM 5 & 744B & \mbox{Terminus 2} & \scorepm{52.4}{2.6} \\
Claude Opus 4.5 & -- & OpenHands & \scorepm{51.9}{2.9} \\
Gemini 3 Flash & -- & \mbox{Terminus 2} & \scorepm{51.7}{3.1} \\
GPT-5.1 & -- & \mbox{Terminus 2} & \scorepm{47.6}{2.8} \\
Kimi K2.5 & 1T & \mbox{Terminus 2} & \scorepm{43.2}{2.9} \\
MiniMax M2.5 & 230B & \mbox{Terminus 2} & \scorepm{42.2}{2.6} \\
DeepSeek-V3.2 & 671B & \mbox{Terminus 2} & \scorepm{39.6}{2.8} \\
Qwen 3 Coder & 480B & \mbox{Terminus 2} & \scorepm{23.9}{2.8} \\
\end{tabular*}
\end{minipage}}
\begin{tabular*}{\textwidth}{@{}c@{\extracolsep{\fill}}c@{}}
\toprule
\usebox{\tbresultsbaselinebox}
&
\begin{minipage}[t][\dimexpr\ht\tbresultsbaselinebox+\dp\tbresultsbaselinebox\relax][t]{0.485\textwidth}
\begin{tabular*}{\linewidth}[t]{@{}>{\raggedright\arraybackslash}p{0.56\linewidth}@{\extracolsep{\fill}}>{\centering\arraybackslash}p{0.21\linewidth}>{\centering\arraybackslash}p{0.18\linewidth}@{}}
\multicolumn{3}{c}{\textbf{Specialized}} \\
\cmidrule{1-3}
\textbf{Model} & \textbf{Scaffold} & \textbf{TB 2.0} \\
\midrule
LiteCoder-Terminal-32B-SFT & \mbox{Terminus 2} & \scorepm{18.5}{3.4} \\
TerminalTraj-32B & \mbox{Terminus 2} & \scorepm{22.0}{5.8} \\
OpenThinkerAgent-32B & \mbox{Terminus 2} & \scorepm{26.2}{1.6} \\
Nemotron-Terminal-32B & \mbox{Terminus 2} & \scorepm{27.4}{2.4} \\
Qwen3-32B + SkillSynth & \mbox{Terminus 2} & \scorepm{29.6}{1.6} \\
CLI-Universe-32B (DS) & \mbox{Terminus 2} & 31.2 \\
Terminal-World-32B & \mbox{Terminus 2} & 31.5 \\
\end{tabular*}
\par\vfill
\begin{tabular*}{\linewidth}[b]{@{}>{\raggedright\arraybackslash}p{0.56\linewidth}@{\extracolsep{\fill}}>{\centering\arraybackslash}p{0.21\linewidth}>{\centering\arraybackslash}p{0.18\linewidth}@{}}
\multicolumn{3}{c}{\textbf{Ours}} \\
\cmidrule{1-3}
Qwen3-32B & NexAU & \scorepm{5.6}{0.9} \\
\quad + Terminal-3.6K & NexAU & \scorepm{32.3}{2.4} \\
Qwen3.5-35B-A3B Base & NexAU & \scorepm{22.5}{0.9} \\
\quad + Terminal-3.6K & NexAU & \textbf{\scorepm{52.0}{3.6}} \\
\end{tabular*}
\end{minipage}
\\
\bottomrule
\end{tabular*}
\end{table}

\subsection{Main Results}
Table~\ref{tab:tb-results} presents the results on Terminal-Bench 2.0. Three observations stand out. First, Terminal-3.6K consistently improves models with substantially different initial capabilities: Qwen3-32B increases from $5.6\%\pm0.9$ to $32.3\%\pm2.4$, while Qwen3.5-35B-A3B increases from $22.5\%\pm0.9$ to $52.0\%\pm3.6$ (absolute gains of 26.7 and 29.5 points). This suggests that the synthesized data provides useful terminal interaction patterns that improve agent capabilities rather than compensating for weaknesses of a particular base model, and the low base-model score (5.6\%) confirms that the harness itself does not inflate performance. Second, NexForge-trained Qwen3-32B achieves the highest mean accuracy compared to terminal-specialized baselines, despite NexForge being a general requirement-driven framework rather than a pipeline tailored to terminal tasks. Third, combining Terminal-3.6K with the stronger Qwen3.5 base yields 52.0\%, reaching the performance range of Claude Opus~4.5 (51.9\%) and Gemini~3~Flash (51.7\%) without changing the model architecture.

\subsection{Scaling Analysis}

\begin{figure}[t]
\centering
\includegraphics[width=\textwidth]{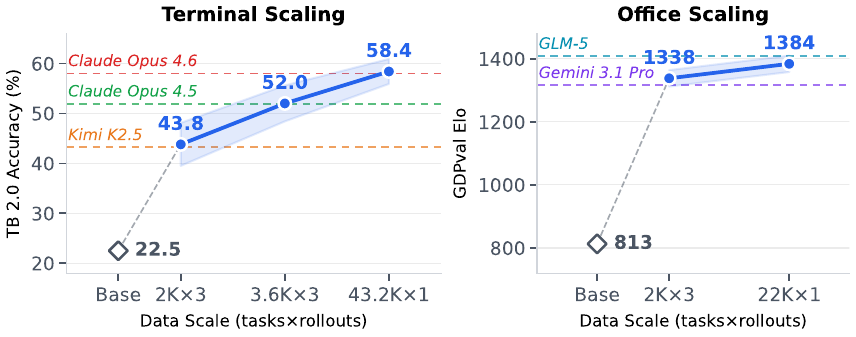}
\caption{\textbf{Data scaling.} Terminal-Bench 2.0 accuracy (left) and GDPval Elo (right) as a function of NexForge data scale on Qwen3.5-35B-A3B.
Dashed lines mark baseline reference models. X-axis labels use the format \emph{tasks${\times}$rollouts}; for example, 2K${\times}$3 denotes 2{,}000 tasks with 3 rollouts each.
}
\label{fig:scaling}
\end{figure}

\paragraph{Domain transfer.}
To demonstrate domain transfer, we replace the terminal requirement with an office-work specification while keeping the pipeline unchanged. NexForge constructs 2{,}000 tasks covering document processing, spreadsheet analysis, planning, communication, and evidence-based recommendations, yielding 5{,}940 trajectories after three rollouts per task. As shown in Table~\ref{tab:gdpval-results}, Office-2K improves Qwen3.5-35B-A3B Base from 813 to 1338 GDPval Elo, comparable to Gemini 3.1 Pro, showing that the same pipeline enables agent post-training for an entirely different capability with only a specification change.

\paragraph{Data volume scaling.}
Within a domain, NexForge's data production is not bottlenecked by curated materials or substrate availability---scaling requires only more tasks. The larger-scale Terminal-43.2K and Office-22K runs use one rollout per task (43{,}200 and 22{,}000 trajectories), as the expanded task pool provides sufficient diversity without repeated rollouts. On Terminal-Bench 2.0, scaling from 3.6K to Terminal-43.2K improves accuracy to 58.4\% (single run) as shown in Figure~\ref{fig:scaling}, confirming that NexForge-generated data effectively scales agent capabilities. The same pattern holds on GDPval: Office-2K (1338) to Office-22K (1384) shows consistent gains (Elo values are contextual as they depend on the comparison pool).

\begin{table}[t]
\caption{GDPval Elo results for office capability evaluation. Elo is calibrated to the GDPval-AA v1 scale~\citep{artificialanalysis2025gdpval} with GPT-5.1 anchored at 1000.}
\label{tab:gdpval-results}
\centering
\small
\setlength{\tabcolsep}{8pt}
\begin{tabular}{@{}llc@{}}
\toprule
Group & Model & GDPval Elo \\
\midrule
\multirow{7}{*}{\textbf{Baselines}}
& GPT-5.4 & 1667 \\
& Claude Sonnet 4.6 & 1633 \\
& GLM-5 & 1408 \\
& Gemini 3.1 Pro & 1316 \\
& MiniMax M2.5 & 1202 \\
& Gemini 3 Flash & 1191 \\
& GPT-5.1 & 1000 \\
\midrule
\multirow{2}{*}{\textbf{Ours}}
& Qwen3.5-35B-A3B Base & 813 \\
& \quad + Office-2K & \textbf{1338} \\
\bottomrule
\end{tabular}
\end{table}

\subsection{Ablation Studies}
\label{sec:ablation}

We first examine whether task-form control and scenario grounding provide complementary signals for task compilation. Using the same 2{,}000 tasks, three rollouts per task, and identical SFT settings, we construct two ablation variants and train them on Qwen3.5-35B-A3B Base to compare with Terminal-2K. \emph{W/o profile} retains the scenarios but removes task-form directives, allowing the agents to infer tasks directly from each scenario. \emph{W/o scenario} retains the task-form profiles but replaces scenarios with generic neutral seeds that provide no contextual grounding.

\begin{figure}[!t]
\centering
\includegraphics[width=\textwidth]{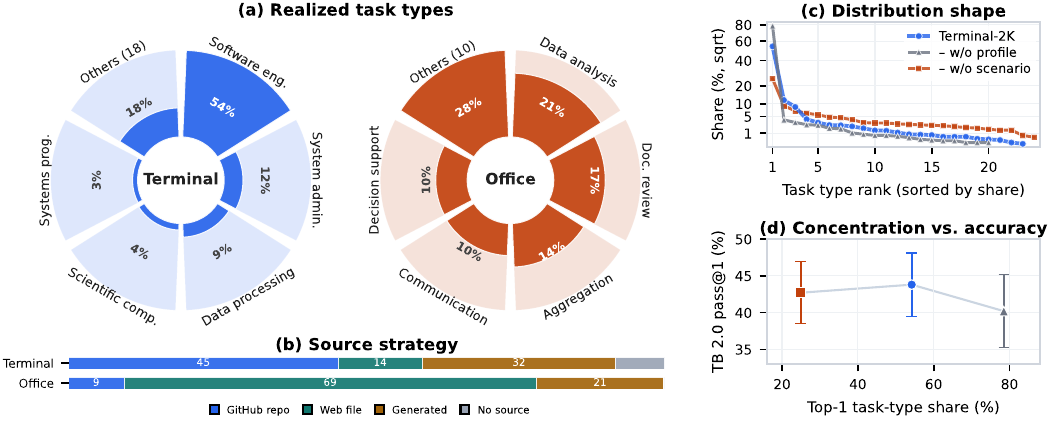}
\caption{\textbf{Distribution analysis.} (a) Task-type frequencies, (b) grounding-source composition, (c) assistant message counts, and (d) tool-call counts per trajectory across ablation variants.}
\label{fig:distributions}
\end{figure}

As shown in Table~\ref{tab:ablation}, the full setting achieves the highest mean accuracy and longest trajectories. Figure~\ref{fig:distributions}(c,d) explains this improvement. We measure task-type diversity by the effective number of types $\exp(H)$, where
$$
H=-\sum_i p_i\ln p_i
$$is the Shannon entropy over task-type proportions; $\exp(H)$ equals the number of equally frequent types yielding the same entropy. Under this measure, \emph{w/o profile} collapses to one dominant task type (78.5\%, $\exp(H)\!\approx\!3$), whereas \emph{w/o scenario} expands coverage to $\exp(H)\!\approx\!16$ while producing an overly flat distribution. Neither meets real-world demand. Terminal-2K maintains $\exp(H)\!\approx\!6$, balancing diversity and frequency. Thus, task-form control prevents mode collapse, while scenario grounding shapes a realistic distribution.

We further study DATC, the compatibility filter that screens task-form options against each scenario during compilation. On average, it retains 4.4 of 24 task types, 5.2 of 8 deliverables, 2.7 of 4 source strategies, and 4.5 of 20 runtimes, leaving only 2.0\% of all combinations compatible. Qwen3.7-Max judges whether each directive composed by GPT-5.5 matches its scenario, comparing DATC with direct sampling (\emph{w/o DATC}). DATC achieves an 81.0\% match rate, versus 7.0\% without it as shown in Table~\ref{tab:datc-quality}, confirming that compatibility modeling is essential for coherent, scenario-aligned directives.

Together, these results reveal three complementary roles: task-form control prevents mode collapse, scenario grounding shapes a realistic distribution, and DATC ensures directive--scenario coherence, each addressing a distinct failure mode in the pipeline.

\begin{table}[t]
\caption{\textbf{Directive compatibility.} LLM-judge match rate between directives and scenarios.}
\label{tab:datc-quality}
\centering
\small
\setlength{\tabcolsep}{12pt}
\begin{tabular}{@{}lccc@{}}
\toprule
\textbf{Corpus} & \textbf{Method} & \textbf{Match count} & \textbf{Match rate} \\
\midrule
\multirow{2}{*}{Terminal-2K}
 & DATC & 1{,}620 / 2{,}000 & 81.0\% \\
 & w/o DATC & 140 / 2{,}000 & 7.0\% \\
\bottomrule
\end{tabular}
\end{table}

\begin{table}[t]
\caption{\textbf{Ablation studies.} Asst./Tool: mean assistant messages and tool calls per trajectory. TB 2.0: pass@1 accuracy on Terminal-Bench 2.0.}
\label{tab:ablation}
\centering
\small
\setlength{\tabcolsep}{6pt}
\begin{tabular}{@{}lccccc@{}}
\toprule
\textbf{Variant} & \textbf{Scenario} & \textbf{Task Form} & \textbf{Asst.} & \textbf{Tool} & \textbf{TB 2.0} \\
\midrule
Terminal-2K    & Reservoir & Profile & 114 & 149 & \textbf{\scorepm{43.8}{4.3}} \\
-- w/o profile  & Reservoir & None & 98  & 133 & \scorepm{40.2}{5.0} \\
-- w/o scenario & Neutral seed & Profile & 95  & 125 & \scorepm{42.7}{4.2} \\
\bottomrule
\end{tabular}
\end{table}

\FloatBarrier

\subsection{Further Analysis}

\paragraph{Cross-domain distribution shift.}
Figure~\ref{fig:distributions}(a,b) shows that NexForge adapts task composition and grounding to user requirements. The terminal corpus is dominated by software engineering (54\%) but covers 23 additional types; the office corpus is more evenly distributed across data analysis, document review, aggregation, and decision support. Grounding differs accordingly: 45\% of terminal tasks use public repositories, while 69\% of office tasks rely on public web files. These distribution shifts arise from the input specifications alone, confirming cross-domain scalability without manual workflow design.

\paragraph{Controlled diversity.}
The scenario reservoir contains 5{,}600 scenarios (combined terminal and office) with no exact or embedding near-duplicates (cosine threshold 0.85). A nearest-centroid classifier on scenario embeddings separates the two domains with 0.98 accuracy, confirming both within-domain diversity and clear cross-domain differentiation in scenario coverage.

\paragraph{Environment quality.}
We manually inspect 100 task packages on five quality dimensions (instruction clarity, material completeness, workspace executability, difficulty calibration, absence of solution leakage), each rated 1--5; most score at least 4 and none below 3. Terminal-2K w/o scenario scores lowest on material completeness due to its lack of scenario grounding. Terminal-2K w/o profile obtains high per-task quality yet still underperforms on Terminal-Bench 2.0, indicating that the ablation gap stems from corpus-level distribution rather than per-package quality differences.

\section{Conclusion and Future Work}

In this work, we present NexForge, a requirement-driven framework that scales agent task synthesis from high-level capability requirements. By decoupling task-form control and scenario grounding from predefined substrates, NexForge transfers across capability domains and scales to larger volumes with only a specification change. Experiments show that Terminal-3.6K consistently improves base models with very different initial capabilities---Qwen3-32B gains 26.7 points and Qwen3.5-35B-A3B gains 29.5 points on Terminal-Bench 2.0---while the same pipeline transfers to office work with only a specification change, reaching 1338 GDPval Elo comparable to Gemini~3.1~Pro. Scaling from 3.6K to 43.2K tasks further improves accuracy to 58.4\%, confirming that NexForge's data production is not bottlenecked by curated materials. NexForge scales to tens of thousands of tasks, enabling the training of the publicly released Nex-N2 model family. Nex-N2-Pro reaches 75.3\% on Terminal-Bench 2.1 and 1585 Elo on GDPval, competitive with frontier proprietary systems including GPT-5.5 and Claude Opus 4.7; Nex-N2-Mini, trained on a smaller data budget, already surpasses the controlled runs, validating that requirement-driven synthesis effectively scales agent capabilities from research-stage corpora to production-grade post-training across terminal, office, and long-horizon agentic reasoning domains. Ablations confirm that task-form control, scenario grounding, and directive--scenario compatibility filtering are each essential for producing coherent, diverse agent tasks aligned with real-world demand. The synthesized tasks currently lack machine-verifiable outcomes, precluding use as benchmarks or RL environments; generating reference solutions and verifier models would close this loop, and extending NexForge to scientific computing and multimodal reasoning would test generalization beyond the current domains.

\bibliographystyle{iclr2026_conference}
\bibliography{references}

\appendix

\section*{Appendix}

This supplementary material documents the implementation details, complete statistics, and additional analyses that support the main paper's claims on scalable agent post-training data synthesis. It is organized as follows. We first give the formal scenario-conditioned task-compilation procedure (Algorithm~\ref{alg:directive-algorithm}) and then walk through a complete end-to-end case study of one synthesized terminal task, from the sampled directive to the runnable package. We next report the synthesis configuration, the two capability requirement specifications that define the pipeline inputs, and corpus-level statistics, audits of scenario-reservoir and task-description diversity, and the full training configuration. We then provide detailed evaluation results and the exact evaluation protocols behind the main-paper scores, including per-task and pass@k analyses and the terminal ablation. Finally, we include faithful English renderings of the four stage prompts and the teacher-rollout runtime prompt that drive the pipeline.

\section{Scenario-Conditioned Compilation Procedure}
\label{sec:supp-algorithm}

Algorithm~\ref{alg:directive-algorithm} details how NexForge composes a task directive $\delta_j$ from the task demand profile $\Phi(I)$ and a single scenario $g_j$ drawn from the reservoir $\mathcal{G}(I)$. The composer processes the ordered task-form dimensions $\{t,d,\sigma,e\}$ (task type, deliverable, source strategy, runtime) one at a time. For each dimension $k$, a compatibility filter $F_\theta$ inspects the requirement, the scenario, and the fields already selected, and returns the subset $C$ of profile options that remain coherent in this context; the realized field $f_{j,k}$ is then sampled from the profile weights restricted to $C$. Because earlier decisions constrain later ones, the directive stays internally consistent (for example, a chosen runtime does not admit an incompatible deliverable), while the profile weights still guide the corpus-level distribution. Language $\ell_j$ and difficulty $h_j$ are drawn last under the batch-level constraints rather than from scenario compatibility, and the retained candidate sets $A$ together with the per-step rationale $R$ are recorded so that downstream stages instantiate the sampled work rather than re-inferring it from whatever materials happen to be available.

\begin{algorithm}[htbp]
\caption{Scenario-conditioned task compilation}
\label{alg:directive-algorithm}
\textbf{Input}: requirement $I$, profile $\Phi(I)$, scenario $g_j$\\
\textbf{Output}: task directive $\delta_j$
\begin{algorithmic}[1]
\STATE $S,A,R\leftarrow\emptyset,\emptyset,\emptyset$
\FOR{$k\in\{t,d,\sigma,e\}$}
\STATE $C\leftarrow F_\theta(I,g_j,k,\Phi_k,S)$
\STATE $A[k]\leftarrow C$
\STATE $f_{j,k}\sim\operatorname{Normalize}(\Phi_k|_{C})$
\STATE $S\leftarrow S\cup\{f_{j,k}\}$
\STATE Append the compatibility rationale to $R$
\ENDFOR
\STATE Sample $\ell_j$ and $h_j$ under the batch constraints
\STATE \textbf{return} $\delta_j=\langle g_j,S,A,\ell_j,h_j,R\rangle$
\end{algorithmic}
\end{algorithm}

\FloatBarrier

\section{End-to-End Case Study}
\label{sec:supp-case-study}

\begin{figure}[t]
\centering
\includegraphics[width=0.6\textwidth]{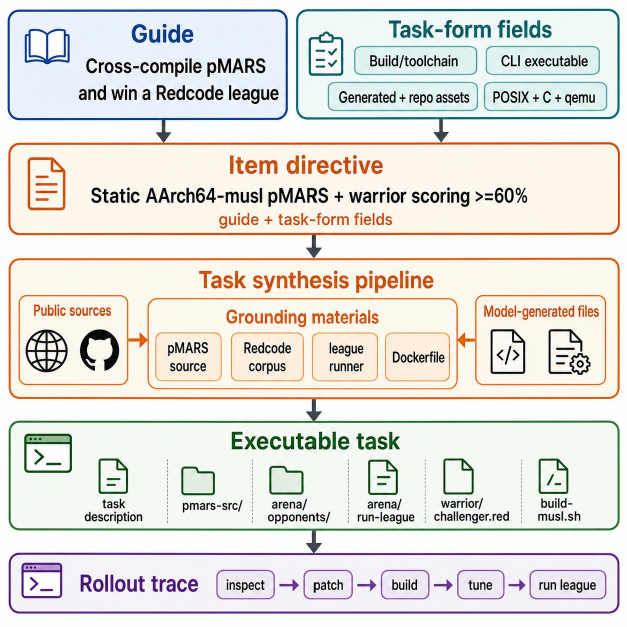}
\caption{End-to-end Terminal-3.6K example: pMARS cross-compilation scenario composed into an executable workspace.}
\label{fig:supp-case-flow}
\end{figure}

To make the pipeline concrete, we trace a single terminal task from the sampled directive to the runnable package (Figure~\ref{fig:supp-case-flow}). The example illustrates the central claim of the paper: the work contract is fixed by the compiled directive \emph{before} any material is gathered, so the substrate is selected to serve the intended task rather than the task being inferred from whatever substrate is convenient.

\paragraph{Scenario and directive.}
The scenario describes a systems-programming context around \emph{pMARS}, the portable Memory Array Redcode Simulator used to run Core War tournaments. Following Algorithm~\ref{alg:directive-algorithm}, the composer conditions the profile on this scenario and sequentially selects compatible task-form fields: a \emph{build, packaging, and toolchain integration} task type, a \emph{command-line executable plus reproducible build script} deliverable, a \emph{GitHub repository adaptation} source strategy, and a \emph{shell/CLI} runtime. Crucially, the same scenario could have supported a build repair, a benchmarking harness, a data-extraction task, or a cross-compilation deliverable; the sampled directive commits to one primary goal---producing an offline, statically linked cross-compilation of pMARS---so later stages do not silently drift toward the most convenient interpretation of the repository.

\paragraph{Material mining.}
Conditioned on the directive, the mining stage performs targeted web research to recover the real workflow and candidate materials: the upstream pMARS source tree, a set of opponent Redcode warriors, and a league runner that scores warriors head-to-head. Mining records each asset with a concrete source, format, and intended role, and distinguishes solver-facing inputs from materials reserved for the runtime environment, so that the downstream package contains genuine artifacts rather than synthetic stand-ins.

\paragraph{Blueprint ideation.}
The ideation stage consolidates the mined evidence into a single-schema blueprint. It fixes the objective (reconstruct a musl-based static pMARS binary offline and validate it by running a warrior through the league), the input materials, the expected deliverables and their success boundary, the workspace layout, and the software dependencies. Because the solver environment is a CPU-only unprivileged container, ideation resolves any environment-sensitive choices at design time---selecting an offline source archive, pinning the toolchain, and avoiding privileged or GPU-dependent steps---so the resulting task is guaranteed to be executable under the standard runtime.

\paragraph{Workspace generation.}
The generation stage instantiates the blueprint into a solver-facing package: it lands the prepared pMARS sources and opponent programs, adds the league runner and any generated fixtures, writes a concise instruction that states the goal, starting materials, and delivery contract without leaking internal checks, and emits a CPU-only, unprivileged Dockerfile that installs the declared dependencies. Automated checks then verify material completeness, source-strategy consistency, runtime restrictions, and the absence of leaked solutions before the package is admitted for teacher rollouts.

\paragraph{Why the directive matters.}
This example shows the difference between requirement-driven and substrate-bound synthesis in miniature. A substrate-bound pipeline that began from the pMARS repository would most naturally emit an implementation or test-repair task, because that is what the repository most directly supports. By fixing a cross-compilation build contract before mining, NexForge instead produces a long-horizon toolchain task whose difficulty and structure are determined by the intended work, and the resulting trajectory is correspondingly deep (many build, inspection, and validation steps rather than a single edit-and-test loop).

\FloatBarrier

\section{Agent Post-Training Data: Synthesis Configuration and Corpus Statistics}

This section documents the configuration that drives synthesis and the corpus-level statistics of the resulting data. We first report the reviewed task-form profile and keyword pools that define the candidate space, then the scale and token/operation composition of the collected trajectories.

\subsection{Reviewed Profile and Keyword Pools}

The reviewed task-form profile defines the candidate space that later sampling draws from. For Terminal-2K it contains 24 primary task types, 20 runtime environments, 8 deliverable types, and 4 source strategies, supported by a domain keyword pool of 3,678 candidates; the office run contains 15 task types, 12 runtimes, 23 deliverables, and 4 source strategies with a 773-keyword pool. The terminal profile spans more task types and runtimes, reflecting the broader operational surface of command-line work, whereas the office profile enumerates more deliverable types, matching its emphasis on documents, analyses, and recommendations. All candidate lists are pruned by the LLM review step for mutual exclusivity and clear semantics before sampling; the realized coverage and concentration of each dimension are reported in Table~\ref{tab:supp-taskform-audit}, and the per-item scene-conditioned narrowing of these pools in Table~\ref{tab:supp-filtering}.

\subsection{Trajectory Scale and Token Mix}

Table~\ref{tab:supp-data} separates original teacher rollouts from the framework-internal cleaned records for each task group and reports median per-rollout token counts alongside mean deduplicated tool calls per trajectory, while Table~\ref{tab:supp-trace-mix} shows how tokens and operations distribute across message roles and tool classes. The terminal corpora are markedly more tool- and build-intensive, whereas the office corpus yields shorter, more read- and document-oriented trajectories.

\begin{table}[tp]
\caption{Trajectory scale per task group. Tasks: distinct tasks yielding at least one valid trace; this count may fall below Packages because some synthesized tasks fail Dockerfile validation, are not executable in the E2B sandbox, or produce no usable trace. Tool/traj.: mean deduplicated tool messages per trajectory, consistent with the main-paper ablation table.}
\label{tab:supp-data}
\centering
\small
\setlength{\tabcolsep}{3pt}
\begin{tabular}{@{}lrrrrrrr@{}}
\toprule
Task group & Packages & Tasks & Rollouts & Records & Tok. B & Med. tok/roll. & Tool/traj. \\
\midrule
Terminal-3.6K & 3,600 & 3,501 & 8,521 & 17,635 & 3.350 & 120.0K & 136 \\
Office-2K & 2,000 & 2,000 & 5,940 & 6,388 & 0.693 & 105.9K & 55 \\
Terminal-2K & 2,000 & 1,947 & 4,656 & 9,214 & 1.760 & 119.3K & 149 \\
w/o profile & 2,000 & 1,916 & 4,574 & 6,555 & 1.019 & 121.7K & 133 \\
w/o scenario & 2,000 & 1,894 & 4,541 & 11,039 & 1.826 & 126.1K & 125 \\
\bottomrule
\end{tabular}
\end{table}

\begin{table}[tp]
\caption{Token and operation mix in converted trajectories.}
\label{tab:supp-trace-mix}
\centering
\small
\setlength{\tabcolsep}{3pt}
\begin{tabular}{@{}lrrrrrrr@{}}
\toprule
Task group & Steps & Tool & Reason B & Tool-call B & Read & Write & Build/test \\
\midrule
Terminal-3.6K & 82.2 & 16.0 & 1.171 & 1.574 & 26.4 & 51.5 & 4.7 \\
Office-2K & 55.6 & 39.7 & 0.072 & 0.313 & 42.8 & 34.1 & 1.0 \\
Terminal-2K & 81.6 & 16.2 & 0.526 & 0.907 & 24.7 & 52.2 & 4.7 \\
w/o profile & 68.3 & 30.0 & 0.257 & 0.435 & 34.8 & 48.3 & 3.8 \\
w/o scenario & 76.3 & 22.1 & 0.552 & 0.837 & 28.5 & 45.1 & 3.2 \\
\bottomrule
\end{tabular}
\end{table}

\subsection{Capability Requirement Specifications}

The two corpora are generated from independently written capability requirements. Each specification is a high-level natural-language description of the target capability that serves as the pipeline input $I$ (Section~\ref{sec:dataset-specifications}). Below are faithful English renderings of the two specifications.

\paragraph{Terminal capability requirement.}
\begin{promptbox}
Design terminal-agent tasks that cover the distribution of multi-domain technical tasks in real command-line environments. The task scope spans software engineering, system administration, security engineering, scientific computing, data science, data processing, data querying, machine learning, model training, mathematical problem solving, optimization, code comprehension, file operations, game or strategy solving, multimedia processing, personal-assistant automation, and other primary directions, with room for further expansion based on real terminal task forms. Each item should require the agent to carry out real technical work through shell commands in a controlled terminal environment and produce deliverables with a well-defined final state, external behavior, or file-level result. Task outputs should be reproducible, executable, and machine-verifiable---for example, source-code or script changes, system configurations or environment states, SQL or structured-data results, model or numerical outputs, optimization plans, processed files or media, game/strategy solutions, automation execution results, build artifacts, or test outcomes. Generated items should vary substantially across task category, runtime environment, input content, required tool chain, operation path, boundary conditions, and success criteria.
\end{promptbox}

\paragraph{Office capability requirement.}
\begin{promptbox}
Design high-value professional work tasks situated in real organizations. The task scope covers digital knowledge work in high-output industries and large organizations, including real estate and leasing, government and public services, manufacturing, professional scientific and technical services, healthcare, finance and insurance, retail and wholesale, information and media, logistics, and operations support. Task roles should cover operations management, asset management, public services, compliance oversight, engineering and manufacturing support, procurement and supply chain, information systems, legal, accounting, project management, healthcare administration, customer service, financial analysis, sales, news editing, audio/video production, and other functions. Each item should require the agent to understand a concrete work scenario, leverage the given information, structured data, communication records, business rules, system exports, public web information, or lightweight tool environments to complete analysis, judgment, computation, planning, coordination, communication, comparison, content production, media processing, or decision support. Task outputs should be concrete deliverables directly usable in a work setting---for example, structured tables, business conclusions, replies to clients or colleagues, process schedules, decision memos, priority judgments, resource-allocation recommendations, calibration notes, updated documents, presentation materials, charts, content drafts, edit lists, subtitle files, media outputs, or audio/video processing results. Generated items should vary substantially across industry, role, objective, information carrier, constraint conditions, collaboration target, time requirements, judgment criteria, and delivery format.
\end{promptbox}

\FloatBarrier
\section{Agent Post-Training: Fine-Tuning Configuration}

This section records the teacher rollout and supervised fine-tuning configuration behind every trained model, so that each reported run can be reproduced from its logged provenance.

Across the experiments, teacher rollouts use DeepSeek V4 Pro.
The records retain all teacher rollouts that survive trace cleaning, which drops only malformed and degenerate trajectories.
The same inclusion rule is used for the terminal ablation and Qwen3-32B comparison.
All SFT runs use 16 nodes with 8 NVIDIA H100 80\,GB GPUs per node, for 128 H100 GPUs in total.
Each node provides 64 CPU cores and 1{,}700\,GiB RAM, interconnected via InfiniBand (NCCL with RDMA).
The software stack comprises Megatron-LM 25.02, CUDA 12.8, and ms-swift for SFT orchestration, running in a containerized Linux environment.

\begin{table}[tp]
\caption{SFT run provenance from training logs.}
\label{tab:supp-sft-runs}
\centering
\small
\setlength{\tabcolsep}{5pt}
\begin{tabular}{@{}lllrrccr@{}}
\toprule
Run & Student & Task set & Train & Val & TP/EP/CP/PP & Time & Ckpt \\
\midrule
Terminal-2K & Qwen3.5-35B-A3B & Terminal-2K both-use & 9,198 & 16 & 2/8/8/1 & 12.2h & 521 \\
w/o profile & Qwen3.5-35B-A3B & Terminal-2K scenario-only & 6,539 & 16 & 2/8/8/1 & 7.4h & 313 \\
w/o scenario & Qwen3.5-35B-A3B & Terminal-2K profile-only & 11,023 & 16 & 2/8/8/1 & 12.7h & 541 \\
Qwen3-32B comp. & Qwen3-32B & Terminal-3.6K & 17,619 & 16 & 2/1/8/4 & 6.5d & 997 \\
\bottomrule
\end{tabular}
\end{table}

All experimental SFT runs use full-parameter SFT with the shared hyperparameters listed in Table~\ref{tab:supp-sft-hparams}; model-specific parallelism configurations are given in Table~\ref{tab:supp-sft-runs}.

\begin{table}[htbp]
\caption{Complete SFT hyperparameters shared across all runs.}
\label{tab:supp-sft-hparams}
\centering
\footnotesize
\setlength{\tabcolsep}{5pt}
\begin{tabular}{@{}ll@{}}
\toprule
Hyperparameter & Value \\
\midrule
Tuner type & Full-parameter \\
Epochs & 5 \\
Max sequence length & 262{,}144 \\
Packing & Enabled (right truncation) \\
Precision & bfloat16 \\
Global batch size & 64 \\
Micro batch size & 1 \\
Learning rate & $10^{-5}$ \\
Minimum learning rate & $10^{-7}$ \\
LR schedule & Cosine decay \\
Warmup fraction & 5\% \\
Weight decay & 0.1 \\
Adam $\beta_1$, $\beta_2$ & 0.9, 0.95 \\
Gradient clipping & 1.0 \\
Seed & 42 \\
Loss scale & \texttt{ignore\_empty\_think} \\
Non-thinking prefix & Added \\
Sequence parallelism & Enabled \\
Attention backend & Flash \\
Recomputation & Full / uniform / 2 layers \\
Cross-entropy fusion & Enabled \\
Val split ratio & 0.001 \\
\midrule
\multicolumn{2}{@{}l}{\textit{MoE-specific (Qwen3.5-35B-A3B only):}} \\
Expert parallelism & 8 \\
Expert capacity factor & 1.5 \\
Auxiliary loss coeff. & $10^{-3}$ \\
Grouped GeMM & Enabled \\
Shared expert overlap & Enabled \\
WD to QK layernorm & Enabled \\
\bottomrule
\end{tabular}
\end{table}

The Qwen3.5-35B-A3B runs use the \texttt{qwen3\_5} chat template with MoE expert parallelism (TP=2, EP=8, CP=8, PP=1); the Qwen3-32B comparison uses the \texttt{qwen3} template with pipeline parallelism (TP=2, CP=4, PP=8).
For the terminal ablations, final training-log summaries give approximately 1.61B packed train tokens for Terminal-2K, 0.97B for Terminal-2K w/o profile, and 1.72B for Terminal-2K w/o scenario.
These values are computed from the logged mean packed sequence length multiplied by the logged packed train-set size.
They differ slightly from the converted-record tokenizer audit in Table~\ref{tab:supp-data}, which sums visible content, reasoning content, and serialized tool calls before training-framework packing and without chat-template overhead.

\FloatBarrier
\section{Detailed Evaluation Results}

Table~\ref{tab:supp-tb} reports the controlled Terminal-Bench 2.0 pass@1 statistics behind the scalar values in the main paper.
All rows use the same in-house NexAU scaffold, task-suite revision, tools, inference settings, and denominator of 89 tasks per run.
Unfinished trials, still-running trials, agent timeouts, verifier timeouts, nonzero exits, and verifier errors are counted as failures.

\begin{table}[tp]
\caption{Controlled Terminal-Bench 2.0 pass@1 accuracy statistics.}
\label{tab:supp-tb}
\centering
\small
\setlength{\tabcolsep}{4pt}
\begin{tabular}{@{}lrrrrrr@{}}
\toprule
Model & Runs & Pass & Mean & 95\% CI & Range \\
\midrule
Qwen3.5-35B-A3B Base & 4 & 80/356 & 22.5 & 0.9 & [21.3, 23.6] \\
Qwen3.5-35B-A3B + Terminal-3.6K & 4 & 185/356 & 52.0 & 3.6 & [48.3, 56.2] \\
Terminal-2K & 4 & 156/356 & 43.8 & 4.3 & [39.3, 49.4] \\
\quad w/o profile & 4 & 143/356 & 40.2 & 5.0 & [33.7, 46.1] \\
\quad w/o scenario & 4 & 152/356 & 42.7 & 4.2 & [37.1, 46.1] \\
Qwen3-32B Base & 4 & 20/356 & 5.6 & 0.9 & [4.4, 7.0] \\
Qwen3-32B + Terminal-3.6K & 4 & 115/356 & 32.3 & 2.4 & [30.3, 36.0] \\
\bottomrule
\end{tabular}
\end{table}

Table~\ref{tab:supp-gdpval} reports available GDPval Elo confidence intervals behind the scalar values in the main paper.
The scoring follows the GDPval-AA v1 methodology of Artificial Analysis: the outputs of GPT-5.1 are fixed as the 1000-point anchor; every pair of models is then compared on all 220 GDPval tasks with an LLM judge that returns a win, loss, or tie for each task; and Elo scores are fit from the full set of pairwise match outcomes.
All rows are fit in one in-house NexAU-based joint comparison pool and should not be interpreted as external GDPval leaderboard scores.

\begin{table}[htbp]
\caption{GDPval Elo estimates from our internal joint comparison pool.}
\label{tab:supp-gdpval}
\centering
\small
\setlength{\tabcolsep}{4pt}
\begin{tabular}{@{}lrr@{}}
\toprule
Model & Elo & 95\% CI \\
\midrule
GPT-5.4 & 1667 & [1639, 1697] \\
Claude Sonnet 4.6 & 1633 & [1605, 1663] \\
GLM-5 & 1408 & [1385, 1433] \\
Qwen3.5-35B-A3B + Office-22K & 1384 & [1358, 1410] \\
Qwen3.5-35B-A3B + Office-2K & 1338 & [1312, 1364] \\
Gemini 3.1 Pro Preview & 1316 & [1291, 1343] \\
MiniMax M2.5 & 1202 & [1175, 1231] \\
Gemini 3 Flash Preview & 1191 & [1164, 1220] \\
GPT-5.1 & 1000 & [1000, 1000] \\
Qwen3.5-35B-A3B Base & 813 & [787, 840] \\
\bottomrule
\end{tabular}
\end{table}

Table~\ref{tab:supp-reference-tb} records the external Terminal-Bench reference values used for context in the main paper.
These reference values come from the public leaderboard and published terminal/agent-data studies \citep{terminalbenchleaderboard2026,fan2026skillsynth,pi2026terminaldata,peng2026litecoderterminal,wu2026terminaltraj,cheng2026terminalworld,raoof2026openthoughtsagent}.

\begin{table}[htbp]
\caption{Reference Terminal-Bench 2.0 scores used in the main paper.}
\label{tab:supp-reference-tb}
\centering
\small
\setlength{\tabcolsep}{5pt}
\begin{tabular}{@{}llcc@{}}
\toprule
Group & Model & Params & TB 2.0 \\
\midrule
External & Gemini 3.1 Pro & -- & \scorepm{80.2}{2.6} \\
 & DeepSeek-V4-Pro & 1.6T & 67.9 \\
 & Claude Opus 4.6 & -- & \scorepm{58.0}{2.9} \\
 & MiniMax M2.7 & 230B & 57.0 \\
 & GPT-5.2 & -- & \scorepm{54.0}{2.9} \\
 & Claude Sonnet 4.6 & -- & \scorepm{53.4}{2.8} \\
 & GLM 5 & 744B & \scorepm{52.4}{2.6} \\
 & Claude Opus 4.5 & -- & \scorepm{51.9}{2.9} \\
 & Gemini 3 Flash & -- & \scorepm{51.7}{3.1} \\
 & GPT-5.1 & -- & \scorepm{47.6}{2.8} \\
 & Kimi K2.5 & 1T & \scorepm{43.2}{2.9} \\
 & MiniMax M2.5 & 230B & \scorepm{42.2}{2.6} \\
 & DeepSeek-V3.2 & 671B & \scorepm{39.6}{2.8} \\
 & Qwen 3 Coder & 480B & \scorepm{23.9}{2.8} \\
\midrule
Qwen3 Based & LiteCoder-Terminal-32B-SFT & 32B & \scorepm{18.5}{3.4} \\
 & TerminalTraj-32B & 32B & \scorepm{22.0}{5.8} \\
 & OpenThinkerAgent-32B & 32B & \scorepm{26.2}{1.6} \\
 & Nemotron-Terminal-32B & 32B & \scorepm{27.4}{2.4} \\
 & Qwen3-32B + SkillSynth & 32B & \scorepm{29.6}{1.6} \\
 & CLI-Universe-32B (DeepSeek) & 32B & 31.2 \\
 & Terminal-World-32B & 32B & 31.5 \\
\bottomrule
\end{tabular}
\end{table}

\subsection{Per-Task Outcomes and Pass@k}
\label{app:tb-pass-at-k}

Beyond mean accuracy, we examine how the four independent runs distribute over individual tasks. Table~\ref{tab:supp-tbtaskdist} reports, for the base model and the Terminal-3.6K model, how many of the 89 tasks are solved in $0,1,\ldots,4$ of the four runs. Training does not merely convert a handful of borderline tasks: the number of tasks that never pass drops from 54 to 29, while the number solved in all four runs rises from 7 to 29. Terminal-3.6K therefore adds robustly solvable tasks rather than occasional lucky passes.

\begin{table}[htbp]
\caption{Per-task pass-count distribution over four runs.}
\label{tab:supp-tbtaskdist}
\centering
\small
\setlength{\tabcolsep}{4pt}
\begin{tabular}{@{}lrrrrr@{}}
\toprule
Model & 0/4 & 1/4 & 2/4 & 3/4 & 4/4 \\
\midrule
Base & 54 & 12 & 8 & 8 & 7 \\
Qwen3.5 + Terminal-3.6K & 29 & 8 & 8 & 15 & 29 \\
\bottomrule
\end{tabular}
\end{table}

\begin{figure}[htbp]
\centering
\includegraphics[width=0.75\textwidth]{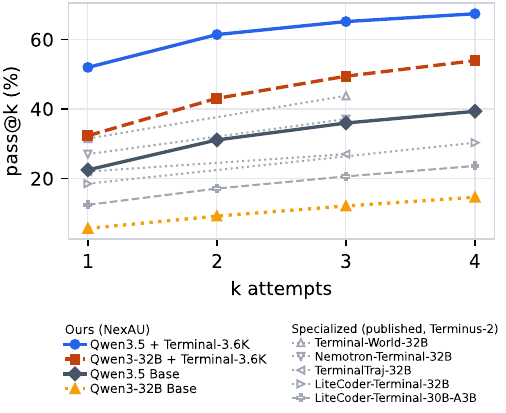}
\caption{Task-level pass@k on Terminal-Bench 2.0.}
\label{fig:tb-pass-at-k}
\end{figure}

Figure~\ref{fig:tb-pass-at-k} extends this to task-level pass@k, estimating whether at least one of $k$ sampled attempts solves each task. NexForge improves the entire pass@k curve for both base models: Qwen3.5-35B-A3B rises from 52.0\% at pass@1 to approximately 68\% at pass@4 and stays roughly 30 points above its base throughout, while Qwen3-32B rises from 32.3\% to approximately 55\% (its base reaches only about 14\% at four attempts). The NexForge-trained Qwen3-32B also stays above the published terminal-specialized models across the evaluated sampling budgets. These trends indicate that Terminal-3.6K expands the set of tasks for which a valid solution can be discovered through repeated sampling, not only single-attempt accuracy; the specialized-model curves are contextual references evaluated with a different scaffold.

\FloatBarrier

\section{Terminal Ablation Details}

The terminal ablation keeps the student, rollout budget, SFT recipe, and Terminal-Bench 2.0 target fixed while changing the synthesis controls.
The Terminal-2K w/o profile variant keeps the diverse scenario reservoir but removes selected task form fields from the directive interface.
In Terminal-2K w/o profile, downstream stages receive the scenario but not selected task form dimensions.
The Terminal-2K w/o scenario variant keeps sampled task form fields and replaces the concrete scenario with a neutral scenario seed.
In the implementation, Terminal-2K w/o scenario samples directly from the reviewed profile without scenario-conditioned filtering and uses the same neutral scenario text for all items.
The Terminal-2K control is the original scenario-conditioned directive construction.

\begin{table}[htbp]
\caption{Primary task type distribution for the terminal ablation (\%).}
\label{tab:supp-ablation-tasktypes}
\centering
\small
\setlength{\tabcolsep}{3pt}
\begin{tabular}{@{}lrrrr@{}}
\toprule
Variant & Labeled & Software & Sys. admin & Data proc. \\
\midrule
Terminal-2K & 2000 & 54.3 & 11.8 & 8.6 \\
w/o profile & 1997 & 78.5 & 0.5 & 2.7 \\
w/o scenario & 2000 & 24.9 & 9.1 & 5.5 \\
\bottomrule
\end{tabular}
\end{table}

Table~\ref{tab:supp-ablation-tasktypes} shows the strongest distributional effect.
w/o profile collapses toward software engineering tasks and almost eliminates some terminal task families, consistent with the model's default mapping from terminal scenarios to tasks.
w/o scenario is much flatter because task form fields are sampled from the reviewed profile without a concrete scenario filter.
At the reconstructed construction-dimension level, w/o scenario has normalized required-task type entropy 0.875 and effective support 16.15, compared with 0.565/6.02 for w/o profile and 0.571/6.13 for Terminal-2K.
These effective-support values count distinct dimension candidates available before generation; they are not the realized task-type distribution in Table~\ref{tab:supp-ablation-tasktypes}, which is far more concentrated for w/o profile (post-hoc labeling collapses to an effective count near one) because the generator defaults to software-engineering tasks when the directive removes selected task types.
w/o profile and Terminal-2K therefore share almost the same construction-level support, yet differ sharply in realized diversity and downstream accuracy.
This explains why w/o scenario remains competitive at 2,000 tasks, while the scenario reservoir is still important for large-scale synthesis where repeated neutral scenarios would make concrete task contexts generic.

\FloatBarrier

\section{Task-Package Audits for Agent Training Data Quality}

\begin{figure}[htbp]
\centering
\includegraphics[width=0.75\textwidth]{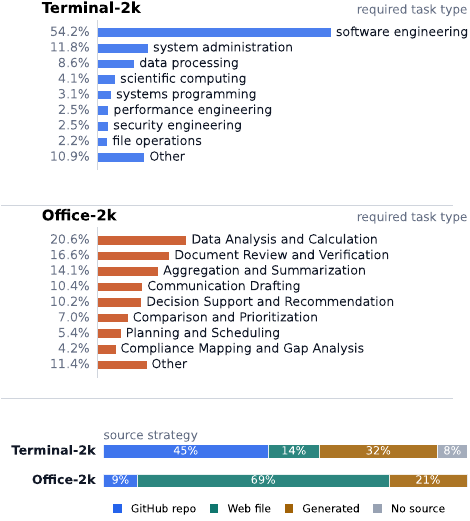}
\caption{Task type and source-strategy distributions for Terminal-2K and Office-2K.}
\label{fig:supp-task-type-distribution}
\end{figure}

We audit the synthesized packages along three axes: static integrity, scenario-conditioned filtering, and the realized distribution over task-form dimensions. Table~\ref{tab:supp-audit} checks that each package contains the expected artifacts and is grounded in real sources; Table~\ref{tab:supp-filtering} reports how aggressively the scenario-conditioned filter narrows the candidate pool before sampling; and Table~\ref{tab:supp-taskform-audit} with Figure~\ref{fig:supp-task-type-distribution} show the realized coverage and concentration of each dimension. Terminal-1.6K is an intermediate corpus produced during the filtering-table audit; it uses a narrower profile (18 rather than 24 primary task types) but otherwise follows the same synthesis pipeline.

\begin{table}[htbp]
\caption{Static synthesis audit.}
\label{tab:supp-audit}
\centering
\small
\setlength{\tabcolsep}{3pt}
\begin{tabular}{@{}lrrrr@{}}
\toprule
Task group & Tasks & Scenarios & Core files & Real src. \\
\midrule
Terminal-3.6K & 3,600 & 3,600 & 3,600 & 2,292 \\
Office-2K & 2,000 & 2,000 & 2,000 & 1,569 \\
\bottomrule
\end{tabular}
\end{table}

\subsection{Task Quality Audit}

To confirm that the synthesized packages are coherent and feasible, we manually inspect stratified task packages sampled from each corpus and ablation variant on a rubric covering five dimensions: description clarity (D1), material completeness (D2), workspace executability (D3), difficulty calibration (D4), and absence of internal-artifact leakage (D5). Overall quality is high: the majority of inspected packages score at least 4 on all five dimensions, and none receives any score below 3. Difficulty calibration is the most frequent deduction---a few tasks pack more work than the stated hour budget---while no internal-artifact leakage is observed. The w/o scenario ablation obtains the lowest per-task quality---particularly on material completeness---because removing the scenario signal causes the pipeline to generate ``from-scratch'' tasks with fewer workspace materials; yet w/o scenario still outperforms w/o profile on Terminal-Bench 2.0, confirming that the ablation gap is driven by task-distribution shape rather than per-task package quality.

\begin{table}[htbp]
\caption{Scenario-conditioned compatibility filtering.}
\label{tab:supp-filtering}
\centering
\small
\setlength{\tabcolsep}{3pt}
\begin{tabular}{@{}lrr@{}}
\toprule
Task group & Type pool$\to$kept & File pool$\to$kept \\
\midrule
Terminal-2K & 24$\to$4.42 & 167$\to$27.37 \\
Terminal-1.6K & 18$\to$4.04 & 175$\to$37.78 \\
Office-2K & 15$\to$6.63 & 105$\to$21.61 \\
\bottomrule
\end{tabular}
\end{table}

\begin{table}[tp]
\caption{Realized distribution across task-form and material dimensions.}
\label{tab:supp-taskform-audit}
\centering
\small
\begin{tabular*}{\textwidth}{@{}ll@{\extracolsep{\fill}}cc@{}}
\toprule
Dimension & Control role & Terminal-2K & Office-2K \\
\midrule
Task type & Primary work form & 24/24; top 27.9\% & 15/15; top 17.3\% \\
Deliverable & Output contract & 8/8; top 19.1\% & 23/23; top 11.9\% \\
Source strategy & Grounding choice & 4; GitHub 45.2\% & 4; web 69.2\% \\
Runtime & Execution medium & 20/20; top 22.4\% & 12/12; top 15.7\% \\
Info.\ carrier & Evidence form & 8/8; top 17.1\% & 19/19; top 14.3\% \\
Material file type & File variety & 167/167; top 3.6\% & 97/105; top 4.6\% \\
\bottomrule
\end{tabular*}
\end{table}

\FloatBarrier
\section{Scenario Reservoir Audit}

We audit scenarios with the same embedding model used by the online novelty filter.
The reservoir contains 5{,}600 accepted scenarios across four generation branches (Table~\ref{tab:supp-scenario-branches}), with zero exact text duplicates and zero embedding near-duplicates above the cosine 0.85 threshold (global maximum nearest-neighbor cosine 0.837).
All statistics are computed over the accepted scenario texts before scenario mining and task form composition.
Pairwise cosine is the mean cosine over all unordered scenario pairs in a branch.
Nearest-neighbor (NN) median measures local redundancy within the branch.
The target silhouette uses cosine distance and the two target groups, terminal and office; higher values indicate stronger separation by target.
Cross-target NN is the share of scenarios whose nearest neighbor inside the same branch belongs to the other target group.

\begin{table}[htbp]
\caption{Branch-level embedding audit for the 5{,}600 accepted scenarios (1{,}868 + 1{,}868 + 932 + 932). No exact or embedding near-duplicates are found above the 0.85 cosine threshold.}
\label{tab:supp-scenario-branches}
\centering
\small
\setlength{\tabcolsep}{3pt}
\begin{tabular}{@{}lrrrrr@{}}
\toprule
Branch & Scenarios & Pair & NN & Silh. & Cross NN \\
\midrule
Self-instruct & 1,868 & 0.278 & 0.678 & 0.141 & 0.16\% \\
Knowledge graph & 1,868 & 0.314 & 0.655 & 0.071 & 3.64\% \\
Keyword self-inst. & 932 & 0.295 & 0.625 & 0.098 & 0.86\% \\
Keyword research & 932 & 0.310 & 0.622 & 0.113 & 0.32\% \\
\bottomrule
\end{tabular}
\end{table}

The table supports three observations used in the main text.
First, self-instruct is the most target-conditioned branch: it has the highest target silhouette and the lowest cross-target nearest-neighbor rate.
Second, keyword research and keyword self-instruct are close in local compactness, but keyword research separates the two target intents slightly more strongly.
Third, knowledge graph expansion has the weakest target separation, which is consistent with the branch exploring reusable entities, workflows, and contexts that can appear in both terminal and office settings.

\FloatBarrier
\section{Task-Description Embedding}

\begin{figure}[htbp]
\centering
\includegraphics[width=0.75\textwidth]{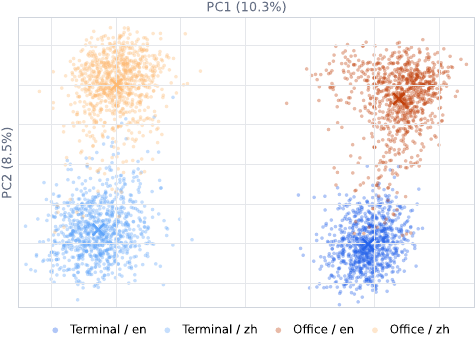}
\caption{PCA projection of task description embeddings for Terminal-2K and Office-2K.}
\label{fig:supp-embedding}
\end{figure}

We embed the final solver-facing task descriptions of the matched Terminal-2K and Office-2K sets (4{,}000 descriptions, 2{,}000 per target, 50/50 English--Chinese) with \texttt{text-embedding-3-small} and analyze their geometry.
The two targets separate cleanly: a nearest-centroid classifier reaches 0.980 accuracy (0.980 for the four target--language groups), and 97.8\% of nearest neighbors share the target (97.2\% share target and language).
Separation is not template collapse, however: within-target random pairs have mean cosine 0.571 versus 0.503 across targets, a modest gap, and nearest-neighbor cosine has median 0.797 with maximum 0.910, so descriptions are semantically concentrated by target and language without duplicating one another.
Figure~\ref{fig:supp-embedding} shows the PCA projection, with four visible target--language clusters.

\FloatBarrier
\section{Synthesis Prompt Templates}

The synthesis pipeline is driven by four stage prompts: \emph{diverse} (task profile), \emph{mine} (scenario research), \emph{ideate} (task blueprint), and \emph{gen} (task materialization); teacher rollouts use the NexAU runtime prompt.
The stage prompts are written in Chinese in our implementation; below are faithful English renderings.
Jinja variables are shown as \texttt{\{\{name\}\}}, and the ablation-mode conditionals (\texttt{guide\_only}, \texttt{intent\_only}) are collapsed to the default both-use path; where a branch changes wording we note it inline.
The verbatim original templates are included in the prompt archive under \texttt{analysis/prompts/}.

\subsection{Planning: Task Profile (diverse)}
This prompt turns a free-form task-set specification into the reviewed task-form profile $\Phi(I)$.

\begin{promptbox}
Carry out task-planning-oriented research based on the user input and organize it into a structured task profile.

User input:
{{ root_query }}

First research and read relevant material around the user input, covering multiple reliable sources. Prefer broad taxonomies, task collections, label systems, tool/ecosystem catalogs, standards, or domain surveys to build the overall frame, then fill long-tail directions with finer material. Scope the research to the domains, workflows, tools, objects, data forms, common deliverables, and runtime environments the user input naturally points to.

The goal is a comprehensive, cleanly structured profile, not a summary and not concrete items. The lists are the reusable candidate space for later generation: cover every candidate with real discriminative value while avoiding semantic overlap within a list. In particular, `task_types` is later sampled by weight for the primary task type, so it must stay same-level, mutually exclusive, and clearly bounded; cross-cutting capabilities, material forms, runtime environments, and delivery forms go in their own fields, and `task_types` keeps only the primary work type.

Output requirements:
- task_types: mutually exclusive primary work categories, i.e. the single kind of work the solver mainly does in one item. Not a scene, exception, material type, runtime, deliverable, verification, or difficulty source. Each item has name, description, when_applicable, and a positive weight (relative sampling weight in the real distribution, need not be normalized; spread head/mid/tail weights apart). Merge items that usually co-occur.
- runtime_environments: runnable, verifiable, interactive environments the task package or Dockerfile must prepare, not a business location. Each item has name, description, tools_or_services, when_applicable (e.g. shell/CLI, Python data analysis, Node/Java/Go/Rust build-test, local web service, database service, browser automation, PDF/OCR, media processing, scientific computing, container tooling, offline document review).
- deliverables: abstract deliverable types the solver should ultimately produce. Each has name, description, typical_success_criteria (an internal profile field summarizing the final quality boundary, NOT a solver-facing acceptance/verification procedure).
- information_carriers: carrier forms of the input information (not deliverables, not runtimes). Each has name, description, when_needed (e.g. code, logs, config, DB exports, issues, email, screenshots, API docs, contracts, policies, media assets).
- material_file_types: a flat list of lowercase suffixes / common format names only (e.g. pdf, csv, xlsx, json, yaml, log, ipynb, pcap). No grouping, no descriptions.
- When the input points to command line / terminal / local repo / executable terminal work, task_types and deliverables should favor runnable, editable, testable, reproducible, machine-checkable work and products.
- Remove near-synonymous items in each list; keep task_types more restrained and mutually exclusive than the others. Keep abstract categories abstract and file suffixes / tool environments concrete.

The final action must call GenerateIntentProfile with output_path={{ output_path }}, providing an intent_profile object containing all of: task_types, runtime_environments, deliverables, information_carriers, material_file_types.
\end{promptbox}

\subsection{Mining: Scenario Research (mine)}
Conditioned on the item directive (a scenario composed with compatible task-form fields), this prompt gathers realistic workflows, terminology, resources, and candidate materials.

\begin{promptbox}
Do focused research around the given scene, distilling the task form, runtime environment, required knowledge, usable information carriers, and delivery boundary that fit this item, and produce a research record for blueprint design.

Scene input (may contain `item_directive`):
{{ seed_scene_json }}

Research focus:
- The work the solver must actually do in the scene, its key objects and constraints.
- How `item_directive.selected` (task type, runtime, deliverable, source type) shapes the task; `required_task_type` is a high-level direction, so research should clarify what it can concretely become within the original user intent, not shrink it to a fixed parameter or fixed procedure.
- How `item_directive.applicable_options` (information carriers, material file types) match the scene.
- The domain, tool, and workflow knowledge and the judgment needed to finish the task.
- Real resources, data, code, logs, config, docs, pages, tests, or interfaces that help. If difficulty_hours is present, treat it as a lower bound on expert time and look for real complexity, edge cases, and delivery boundaries that support it.

Proceed by: (1) understanding the task scene (core need, key objects, constraints, target deliverable); (2) progressive retrieval, splitting more specific queries from prior findings, covering domains, tools, environments, interfaces, data forms, deliverables, and quality boundaries; (3) curating usable resources/assets, each with a concrete name, link, format/type, purpose, and relation to the task, judging whether it is substantial enough to support reasoning, operation, construction, comparison, analysis, decision, delivery, or verification.

Produce a research record with four parts: (I) task interpretation; (II) knowledge and information required; (III) execution key points and pitfalls; (IV) usable resources and asset suggestions, marking which are suited to be solver-facing inputs and which to hidden evaluation, environment dependencies, or background. Output natural-language text with accessible links.
\end{promptbox}

\subsection{Ideation: Task Blueprint (ideate)}
This prompt turns the directive and mined materials into a single-schema blueprint (\texttt{ideation.json}), preparing real sources via \texttt{DownloadSourceFile} / \texttt{PrepareGitHubRepo}. A variant (\emph{history-to-blueprint}) emits the same schema from the recorded research history as an XML-like document.

\begin{promptbox}
You are a professional task-blueprint designer. Based on the input query, scene research, task profile, and item directive, design a high-quality, high-difficulty, executable task blueprint.

Inputs: {{ scene_query }} ; scene report {{ scene_report }} ; task profile {{ intent_profile_json }} ; item directive {{ item_directive_json }}. Target language: {{ target_language_label }}. Difficulty: an expert should need at least {{ difficulty_hours }} hours (a lower bound, not a target to shrink to).

Requirements:
- Understand what the query wants the solver to do and the real-world scene behind it. Do not merely rewrite the guide into a task description; extract scene, role, objects, boundary conditions, final target state, and material forms, then expand into deep real work within the user-intent boundary.
- Choose the task type, runtime, deliverable, source type, and asset form fitting the scene. Runtime means the internal execution environment the package/Dockerfile prepares, not a business location and not a setting to write into the prompt. If `item_directive.selected.required_*` fields exist they set the high-level type/deliverable/runtime/source and must be honored; optional_* are candidates, not an output checklist; default to a single primary task type.
- The final task must have one clear primary goal and one primary task type; other types or deliverables only appear as supporting dimensions. Do not stack several medium tasks or independent products to fake difficulty.
- The solver environment is a CPU-only, ordinary unprivileged container. Do not plan tasks needing GPU/CUDA/NVIDIA/TPU or host-level daemons, Docker Engine/socket, docker compose, KVM, PID 1 systemd, kernel modules, loop mount, or privileged containers; container/OCI directions become file-level checks/fixes on Dockerfile/Containerfile, manifest, tar, or OCI layout. These execution facts are an internal design constraint and must NOT be written into task_description unless the task object itself is a container/OCI artifact.
- Use WebSearch / WebFetch to add background and to look for publicly accessible, appropriately sized real files or GitHub repositories matching the direction. Record the final source choice in `source_selection` (prepared vs to-be-generated assets), referencing canonical tool outputs (remote_rel_path). Source priority: github_repo_adaptation > internet_file_download > generated_files > no_prerequisite_files; use generated files mainly to fill context, fixtures, edge samples, and consistency material, not to replace real data when available.
- The initial solver-facing package must contain only real files and directories (no symlinks); do not materialize /proc, /sys, /dev, snapshots, infinite link trees, full .git history, or large vendor/build output.
- task_description is a concise, direct, solver-facing statement (goal, background, starting materials or clean-environment contract, constraints, deliverable, paths, final state, external behavior contract). Its first sentence starts from the most distinctive information. Do not prescribe the solver's solution order, planning, debugging route, or verification steps; do not expose hidden tests, reference answers, expected value ranges, scoring details, injected bugs, or a known-root-cause list. Keep local paths relative.
- Do not mention internal artifacts (blueprint, ideation, item directive, intent profile, source_selection, prepared/generated assets, manifest, research process, tool names, hidden checks, the generation stage, or "intentionally injected/constructed/simulated" meta-information) in task_description.
- Derive difficulty on the spot (hard_topics, difficulty_budget.justification) from the chosen fields, materials, deliverable, sources, and difficulty_hours; do not use a preset difficulty list.

After research, source preparation, field design, path/environment checks, and key value/contract self-checks, call GenerateIdeationBlueprint with output_path={{ output_path }} (a heavy final tool that regenerates the whole blueprint from the conversation; make small fixes by editing ideation.json directly). Blueprint schema requirements: {{ schema_requirements }}.
\end{promptbox}

\subsection{Generation: Task Materialization (gen)}
This prompt writes the solver-facing package (task description, materials, Dockerfile, and internal audit files) from the blueprint, using prepared sources.

\begin{promptbox}
Synthesize a real task from the task blueprint. You only synthesize; an independent solver will complete it. The blueprint's task_description is a draft to align paths, add detail, and ground materials on. Prepared external files from ideate are used directly.

Blueprint: {{ ideation_json }}. Prepared sources dir {{ prepared_sources_dir }} and manifest {{ prepared_sources_manifest_path }} (read it before handling source_selection.prepared_assets). Target language: {{ target_language_label }}. Generate the Dockerfile at {{ dockerfile_output_path }} from environment_plan.

source_selection handling: prepared_assets are already prepared by ideate; locate them via the manifest (path = remote_rel_path) and copy/trim/extend/extract them into the final package. Source priority github_repo_adaptation > internet_file_download > generated_files > no_prerequisite_files; real sources should be the core material, generated assets only fill context/fixtures/edge samples. Record used assets in generation_summary.json (name, source_type, source_path, copied_to/adapted_to, usage).

Solver-facing filesystem constraints: only real ordinary files and directories; no symlink of any kind; do not copy upstream symlinks, /proc, /sys, /dev, virtual filesystems, snapshots, infinite link trees, full .git history, dependency/build caches, or large vendor/build output; trim to the minimal real subset the task needs.

Internal execution constraints: CPU-only (no GPU/CUDA/NVIDIA/TPU in tasks, Dockerfile, scripts, tests, or data pipelines; shrink ML/image/video/scientific work to small models, small data, CPU inference/training, or pure-algorithm substitutes). Ordinary unprivileged container (no host daemon, Docker Engine/socket, docker compose, KVM, PID 1 systemd, kernel modules, loop mount, or privileged container; container/OCI work is file-level only). Do not write these execution facts into the task description unless the task object is itself a container/OCI artifact.

Landing principles: task_type is a high-level direction, not a fixed script; keep the blueprint intent, source choice, and delivery boundary; deepen one primary goal rather than stacking medium subtasks. The final task_description reads like a concise issue or benchmark instruction (starting materials or clean-environment contract, goal, constraints, deliverable, key interfaces/output schema, final state, external behavior contract); no numbered subtask checklist, no solver solution route, no hidden tests / full verification commands / reference answers / expected value ranges / scoring internals, no known-defect or preset-root-cause list. For fix/recovery/troubleshooting tasks the starting materials may contain failing tests, logs, half-done implementations, or inconsistent data, but only visible symptoms and contracts go into the description.

Deliverables: (1) the task description file at {{ task_description_path }} (relative paths only); (2) reference material files (reuse prepared real sources first, then create real supplementary files that are readable, parseable, and actually used in solving); (3) the Dockerfile at {{ dockerfile_output_path }} (base_image matches the blueprint, installs declared dependencies, CPU-only, unprivileged-compatible, does not COPY the task files in). generation_summary.json and generation_validation.json are internal audit files at the generation root, never inside the solver package and never referenced by the task description.

Workflow: understand the blueprint; read the prepared-sources manifest; land prepared assets, then generated assets and other referenced files; write the Dockerfile; align task_description paths; self-check goal/material/interface consistency and output directory; write generation_validation.json. Finish with a checklist confirming files exist, hard_topics are reflected, materials are real, prepared assets are used and logged, paths are relative and consistent, no leaked internal/execution facts, no symlinks, CPU-only, unprivileged-compatible, and difficulty >= {{ difficulty_hours }} hours.
\end{promptbox}

\subsection{Teacher Rollout Runtime (distillation)}
Teacher rollouts that produce the distillation trajectories run under the NexAU runtime prompt (already in English). The base system prompt and the context-compaction prompt are shown below; tool descriptions are provided per run.

\begin{promptbox}
You are an AI agent named '{{ agent_name }}' built on the NexAU framework.

You have access to the following tools:
{% for tool in tools %}- {{ tool.name }}: {{ tool.description }}
{% endfor %}
You can delegate tasks to the following sub-agents:
{% for sub_agent in sub_agents %}- {{ sub_agent.name }}: {{ sub_agent.description }}
{% endfor %}

Your goal is to help users accomplish their tasks efficiently by:
1. Understanding the user's request
2. Determining if you can handle it with your available tools
3. Delegating to appropriate sub-agents when their specialized capabilities are needed
4. Executing the necessary actions and providing clear, helpful responses
\end{promptbox}

\begin{promptbox}
You have been working on the task described above but have not yet completed it. Write a continuation summary that will allow you (or another instance of yourself) to resume work efficiently in a future context window where the conversation history will be replaced with this summary. Include: (1) Task Overview -- the core request, success criteria, and constraints; (2) Current State -- what is completed, files created/modified/analyzed with paths, key outputs; (3) Important Discoveries -- technical constraints, decisions and rationale, errors and resolutions, approaches that failed and why; (4) Next Steps -- specific remaining actions, blockers, open questions, priority order; (5) Context to Preserve -- user preferences, non-obvious domain details, promises made. Be concise but complete; err on the side of preventing duplicate work or repeated mistakes.
\end{promptbox}

\end{document}